\documentclass[usenatbib]{mnras}
\usepackage{graphicx}
\usepackage{color}
\usepackage{setspace}
\usepackage{amsmath,amssymb,amstext}
\usepackage{dcolumn}
\newcolumntype{d}[1]{D{.}{.}{#1}}

\makeatletter
\newlength{\abovecaptionskip}%
\makeatother
\usepackage{threeparttable}

\usepackage{multicol}
\usepackage{pdflscape}

\def\apj{\rm ApJ}
\def\apjl{\rm ApJL}
\def\apjs{\rm ApJS}
\def\aj{\rm AJ}
\def\jcap{\rm JCAP}
\def\mnras{\rm MNRAS}

\def\aap{\rm AAP}

\def\prl{\rm PRL}
\def\prd{\rm PRD}

\def\lya{Ly$\alpha$}
\def\hmpc{$h^{-1} \mbox{ Mpc}$}

\begin{document}

\title [3PCF of the Lyman-alpha Forest]{UV Background Fluctuations and Three-Point Correlations in the Large Scale Clustering of the Lyman-alpha Forest}
\author[Tie et al. ]{Suk Sien Tie$^1$,
    David ~H. Weinberg$^{1,2}$,
    Paul Martini$^{1,2}$, 
    Wei Zhu$^{3}$,
    S\'ebastien Peirani$^{4,5}$, \newauthor
    Teresita Suarez$^{6,7}$,
    St\'ephane Colombi$^{4}$
    \\
  $^{1}$ Department of Astronomy, The Ohio State University, 140 West 18th Avenue, Columbus OH 43210, USA \\
  $^{2}$ Center for Cosmology and AstroParticle Physics, The Ohio State University, 191 W. Woodruff Avenue, Columbus OH 43210, USA \\
  $^{3}$ Canadian Institute for Theoretical Astrophysics (CITA), University of Toronto, 60 St. George Street, Toronto, ON M5S 3H8, Canada\\
  $^{4}$ Institut d'Astrophysique de Paris, CNRS \& UPMC, UMR 7095, 98 bis Boulevard Arago, 75014, Paris, France\\
  $^{5}$ Universit\'e C\^ote d'Azur, Observatoire de la C\^ote d'Azur, CNRS, Laboratoire Lagrange, France\\
  $^{6}$ Department of Physics and Astronomy, University College London, Gower Street, London WC1E 6BT, UK\\
  $^{7}$ Institute for Astronomy, University of Edinburgh, Blackford Hill, Edinburgh EH9 3HJ, UK
   }

\maketitle
\begin{abstract}
Using the \lya\ mass assignment scheme (LyMAS), we make theoretical predictions for the 3-dimensional 3-point correlation function (3PCF) of the \lya\ forest at redshift $z=2.3$. We bootstrap results from the (100 \hmpc )$^3$ Horizon hydrodynamic simulation to a (1 $h^{-1}$ Gpc)$^3$ $N$-body simulation, considering both a uniform UV background (UVB) and a fluctuating UVB sourced by quasars with a comoving $n_q \approx 10^{-5}$ $h^3$ Mpc$^{-3}$ placed either in massive halos or randomly. On scales of $10-30$ \hmpc , the flux 3PCF displays hierarchical scaling with the square of the 2PCF, but with an unusual value of $Q \equiv \zeta_{123}/(\xi_{12} \xi_{13} + \xi_{12} \xi_{23} + \xi_{13} \xi_{23}) \approx -4.5$ that reflects the low bias of the \lya\ forest and the anti-correlation between mass density and transmitted flux. For halo-based quasars and an ionizing photon mean free path of $\lambda = 300$ \hmpc\ comoving, UVB fluctuations moderately depress the 2PCF and 3PCF, with cancelling effects on $Q$. For $\lambda = 100$ \hmpc\ or 50 \hmpc , UVB fluctuations substantially boost the 2PCF and 3PCF on large scales, shifting the hierarchical ratio to $Q \approx -3$. We scale our simulation results to derive rough estimate of the detectability of the 3PCF in current and future observational data sets for the redshift range $z=2.1 - 2.6$. At $r = 10$ \hmpc\ and 20 \hmpc , we predict a signal-to-noise (SNR) of $\sim$ 9 and $\sim$ 7, respectively, for both BOSS and eBOSS, and $\sim$ 37 and $\sim$ 25 for DESI. At $r = 40$ \hmpc\ the predicted SNR is lower by a factor of $\sim$ 3$-$5. Measuring the flux 3PCF would provide a novel test of the conventional paradigm of the \lya\ forest and help separate the contributions of UVB fluctuations and density fluctuations to \lya\ forest clustering, thereby solidifying its foundation as a tool of precision cosmology.
\end{abstract}

\section{Introduction}
The \lya\ forest arises from the low column density ($N_{\mathrm{HI}} \sim 10^{14} \mbox{ cm}^{-2}$) tenuous gas in mildly overdense regions of the intergalactic medium
(IGM).  Initially thought to stem from discrete gas clouds along the line of sight \citep{Lynds1971,Sargent1980}, a combination of cosmological simulations, analytic models, and improved observations in the mid-1990s established the now
standard view of the \lya\ forest as tracing a smoothly fluctuating and continuous matter distribution \citep{Cen1994,Zhang1995,Hernquist1996,MiraldaEscude1996,Bi1997,Croft1997,Rauch1997}, an inhomogeneous version of the classic Gunn-Peterson effect \citep{GP1965}. In this standard picture, the absorbing gas is in
photoionization equilibrium with the ionizing background radiation, with \lya\
optical depth $\tau = -\ln F \propto n_H^2 T^{-0.7} \Gamma^{-1}$, where $F$ is
the continuum-normalized transmitted flux, $n_H$ is the total hydrogen density,
$T$ is the IGM gas temperature, and $\Gamma$ is the hydrogen photoionization
rate.  The low density gas that fills most of the volume also obeys a power-law
temperature-density relation \citep{Katz1996,HuiGnedin1997} and approximately traces
the underlying dark matter distribution \citep{Croft1999a,Peeples2010}. This allows
a quantitative connection between the \lya\ forest and the dark matter density
field known as the fluctuating Gunn-Peterson approximation (FGPA,
\citealt{Weinberg1998}).

This picture, together with improving cosmological simulations and observational data sets, has turned the \lya\ forest into a powerful probe of matter clustering at redshifts $z=2-4$.  Early cosmological studies focused on the line-of-sight power spectrum or the one-point probability distribution function (PDF) of the transmitted flux \citep{Croft1998,Croft1999a,McDonald2000,Croft2002}, with a large leap in precision enabled by the enormous sample of quasar spectra from the Sloan
Digital Sky Survey (SDSS, \citealt{McDonald2005,McDonald2006}). The Baryon
Oscillation Spectroscopic Survey (BOSS, \citealt{Dawson2013}) of SDSS-III \citep{Eisenstein2011} transformed \lya\ forest cosmology by providing a dense enough grid of sight-lines to enable measurements of 3-d flux auto-correlation functions across sight-lines \citep{Slosar2011} and precise measurements of
cross-correlations between the \lya\ forest and other tracers such as damped-\lya\ systems and quasars
\citep{Ribera2012,Ribera2013,Ribera2014}. These 3-d measurements are especially powerful for cosmology because they enable measurements of the distance-redshift relation and the Hubble expansion via baryon acoustic
oscillations \citep{Busca2013,Slosar2013,Delubac2015,Bautista2017,Bourboux2017}. The large and uniform sample of BOSS spectra also enables highly precise measurements of the line-of-sight power spectrum \citep{Nathalie2013} and flux PDF \citep{Lee2015}.  These 1-d statistics from BOSS and from high-resolution spectra probe small scale dark matter physics, neutrino masses, the amplitude of matter correlations, and the thermal state of the IGM (e.g., \citealt{Bolton2008,Viel2013,Bolton2014,Nathalie2015,Rossi2017,Walther2019,Khaire2019b}).

In this paper we present theoretical predictions for the 3-dimensional 3-point correlation function (3PCF), $\zeta(r_{12},r_{13},r_{23})$, of the \lya\ forest at $z = 2.3$.  Here $\zeta \equiv \langle \delta^F_1\delta^F_2\delta^F_3 \rangle$ where $\delta^F = (F-\bar{F})/\bar{F}$ is the fractional deviation of the transmitted flux at three positions, denoted by the subscripts, that form a triangle with side lengths $r_{ij}$. The 3PCF is the Fourier transform of the bispectrum, just as the 2-point correlation function (2PCF), $\xi(r)$, is the Fourier transform of the power spectrum. A volume average of the 3PCF yields the skewness $\langle \delta_S^3 \rangle$ of the smoothed $\delta$ field just as a volume average of the 2PCF yields the variance $\langle \delta_S^2 \rangle$. \cite{Mandelbaum2003} and \cite{Viel2004} presented numerical and analytic predictions and measurements of the line-of-sight 1-d flux bispectrum, and \cite{Zaldarriaga2001} investigated a correlation between large scale fluctuations and small scale power that is also a form of 1-d bispectrum. To our knowledge, however, ours is the first investigation of the 3-dimensional 3-point flux correlations. We carry this out using a modified form of the \lya\ Mass
Association Scheme (LyMAS, \citealt{Peirani2014,Lochhaas2016}), which bootstraps results from high-resolution hydrodynamic simulations onto large cosmological $N$-body volumes. Our study is motivated by the prospect
of measuring 3-point correlations with the large \lya\ forest sample expected from the Dark Energy Spectroscopic Instrument (DESI, \citealt{DESI2016}), which will measure $10^6-10^7$ \lya\ forest spectra over 14,000 deg$^2$, as well as the possibility of first detections with existing data from BOSS and its SDSS-IV successor eBOSS \citep{Dawson2016}.

The Gaussian initial conditions predicted by standard inflationary models have a vanishing 3-point function. However, gravitational instability of Gaussian initial
conditions generates a non-vanishing 3-point function at second order in perturbation theory, with the scaling 
\begin{equation}
\zeta(r_{12},r_{13},r_{23}) = Q\left[
  \xi(r_{12}) \xi(r_{23}) + \xi(r_{23}) \xi(r_{31}) + \xi(r_{31}) \xi(r_{12})
  \right] ~,
\label{eqn:fry} 
\end{equation} 
where $Q$, often referred to as the reduced 3PCF, is a dimensionless quantity of order unity with
moderate dependence on the shape of the matter power spectrum and the shape of the triangle \citep{Fry1984}. The analogous ``hierarchical'' relation for moments of the smoothed matter density field is $\langle \delta_m^3 \rangle =
S_3 \langle \delta_m^2 \rangle^2$ with $S_3 \approx 3Q$ \citep{Juszkiewicz1993}.
A local bias relation $\delta = f(\delta_m)$ between the matter density contrast and that of a tracer field preserves the hierarchical form of equation~(\ref{eqn:fry}) at second order but changes the value of $Q$ and its dependence on triangle shape \citep{Fry_Gaztanaga1993,Fry1994,Juszkiewicz1995}. 

We will show that the 3PCF of the \lya\ forest scales like the square of the 2PCF as in equation~(\ref{eqn:fry}), but with an unusual value of $Q$ that reflects the low bias factor of the forest flux fluctuations \citep{Slosar2011}. In the nonlinear and strong clustering regime (0.1 $\lesssim$ $r$ $\lesssim$10 $h^{-1}$ Mpc), $Q$ for galaxies has been observed to be constant at $\approx 1.3$ with no clear dependence on triangle shape \citep{Peebles1975,Groth1977}, consistent with $N$-body simulations of the matter distribution \citep{Fry1993,Matsubara1994,Scoccimarro1998,ScoccimarroFrieman1999}. On larger scales, observations, simulations, and perturbation theory suggests that galaxies do not strictly show a constant $Q$ but exhibit scale and shape dependence \citep{Jing1998,Scoccimarro1998,Takada2003,McBride2011,Hoffman2018}. The BAO feature has been detected in the 3PCF measurements of BOSS galaxies \citep{Slepian2017}, while other studies focus on the galaxy bispectrum (e.g., \citealp{Tellarini2016,Desjacques2018,Gualdi2019}). 

Spatial fluctuations of the ionizing ultraviolet background (UVB) and the IGM temperature-density
relation can imprint structure on the \lya\ forest in addition to the clustering generated by the density and velocity fields.  Some level of spatial variation
of $\Gamma$ is inevitable because much of the ionizing background at $z=2-4$ comes from relatively rare quasars, and the expected mean free path of ionizing
photons is only $\sim 100-600$ comoving \hmpc\ \citep{Meiksin2004,Worseck2014}. Fluctuations of the temperature-density relation at this redshift could arise from the residual effects of inhomogeneous He II reionization \citep{Lai2006,White2010,McQuinn2011}.  These effects complicate the relation between the \lya\ forest and the underlying matter density, and they
are a source of systematic uncertainty in cosmological interpretation of \lya\ forest clustering.  Diagnostics of ionizing background fluctuations or temperature fluctuations are valuable both as direct probes of these physical processes and to help control the cosmological systematics.

Early studies of the impact of UVB fluctuations on the
forest focused on the column density distribution and correlation function of \lya\ absorption lines \citep{Zuo1992a,Zuo1992b,Fardall1993}. \cite{Croft1999a} and \cite{Gnedin2002} studied the effect of UVB spatial variations on the flux power spectrum and the recovered matter power spectrum and found a negligible effect on small scales but a potential effect on large scales.  By further including the effect of quasar lifetimes, \cite{Croft2004} found that UVB fluctuations weakly suppress the flux power spectrum at small scales.
\cite{Meiksin2004} examined similar effects at redshifts $z > 5$, where fluctuations are large because of the short photon mean free path. Recent analytical studies by \cite{Pontzen2014} and \cite{Gontcho2014} demonstrated the scale-dependence of a UVB fluctuation imprint on the flux power spectrum and the resultant broadband distortion to the correlation function of the forest. \cite{Suarez2017} extended these studies to include the effect of quasar emission geometry. The impact of temperature fluctuations from inhomogeneous He II reionization is less well explored, but effects are expected to be present \citep{Lai2006,White2010,McQuinn2011}.  

In this paper we aim to establish basic theoretical expectations for the flux 3PCF at $z \sim 2.3$ and to investigate how UVB fluctuations affect the flux 2PCF and 3PCF on scales of $\sim 5-50$ $h^{-1}$ Mpc. Because ionizing background fluctuations modulate the \lya\ flux with a field that is non-Gaussian and has a different power spectrum than the underlying density field, their impact on the 3PCF could be distinctive. We find that a fluctuating UVB changes the 2PCF and 3PCF of the \lya\ forest at all scales to give systematically larger values as the UVB becomes more inhomogeneous. A combination of the 2PCF and 3PCF could then allow better separation between UVB fluctuations and other astrophysical and cosmological parameters. We also use our simulations to give an estimate of the achievable signal-to-noise ratio (neglecting observational noise such as photon noise) of a 3PCF measurement for future and current surveys, in which we predicted a 3PCF detection with a SNR of $\sim 7$ for BOSS and eBOSS and $\sim 25$ for DESI within the redshift range $z=2.1 - 2.6$.

In \S 2 we define our notation for the \lya\ forest 2PCF and 3PCF measurements. In \S 3 we explain how we use LyMAS to predict these clustering statistics for a uniform ionizing background and for a fluctuating background sourced by quasars in massive halos or placed at random, with different choices of source volume density and photon mean free path. Section 4 presents our clustering results with uniform and fluctuating UVB and a rough estimate of detectability of the 3PCF. We summarize our findings in \S 5.

\section{Correlation functions}
For measurements of the \lya\ forest at redshift $z$, we define the flux fluctuations for a pixel at redshift-space position $\mathbf{x}$, 

\begin{equation}
\delta^F(\mathbf{x}) \equiv \frac{F(\mathbf{x})}{\bar{F}(z)} - 1, 
\end{equation}

\noindent where $F = e^{-\tau_{\mathrm{Ly}\alpha}}$ is the ratio of the transmitted flux to the quasar continuum and $\bar{F}(z)$ is the mean transmitted flux at redshift $z$. We define the flux 2PCF

\begin{equation}
\xi(\mathbf{r}) = \left\langle \delta^F(\mathbf{x})\delta^F(\mathbf{x} + \mathbf{r}) \right\rangle,
\end{equation}

\noindent where the average is over all available pixel pairs in a sample of sight-lines with redshift-space separation $\mathbf{r}$. In general, the clustering of the \lya\ forest is highly anisotropic due to redshift-space distortions \citep{Slosar2011}. For simplicity, in this paper we will restrict our attention to purely transverse or (in \S3.3) nearly transverse pixel separations, so that $\mathbf{r} = |\mathbf{r}|$ refers to the transverse separation of sight-lines. For a set of sight-lines through a simulation, we measure the 2PCF by considering all pairs of sight-lines with transverse separations $r_{12} \rightarrow r_{12} + \Delta r_{12}$ and computing 
\begin{equation}
\xi(r_{12}) = \langle \delta^F_1 \delta^F_2 \rangle,
\end{equation}

\noindent where the average includes all transverse pixel pairs along all sight-line pairs. 

The formalism for the transverse 3PCF follows similarly to that of the 2PCF:
\begin{equation}
\zeta(r_{12}, r_{13}, \theta) \equiv \left\langle \delta_1 \delta_2 \delta_3 \right\rangle,
\end{equation}
\noindent
where $r_{12}$ is the separation between the first and second line of sight, $r_{13}$ is the separation between the first and third line of sight, and $\theta$ is the angle between the vectors $r_{12}$ and $r_{13}$. The reduced 3PCF, $Q$, can be constructed from the ratio of the 2PCF and 3PCF according to equation~(\ref{eqn:fry}) as
\begin{equation}
Q(r_{12}, r_{13}, \theta) = \frac{\zeta(r_{12}, r_{13}, \theta)}{\xi(r_{12})\xi(r_{13}) + \xi(r_{12})\xi(r_{23}) + \xi(r_{13})\xi(r_{23})}.
\end{equation}

\section{Simulations and method}
\subsection{Predicting Ly$\alpha$ forest correlations with LyMAS}
Accurately modeling the \lya\ forest with hydrodynamic simulations requires resolving the pressure-support scale (Jeans scale) of the diffuse IGM, which is of order $\lambda_J \sim 0.25$ \hmpc\ comoving for a matter overdensity $\delta \approx 10$ (\citealp{Peeples2010}, eq. 2). Predicting the 3PCF on scales accessible to BOSS and DESI requires simulation volumes of $\sim 1$ Gpc$^3$ or  more, and this combination of volume and resolution is impractical with current capabilities. We therefore compute our flux predictions with LyMAS \citep{Peirani2014}, which uses a high-resolution hydrodynamic simulation to compute the conditional PDF, $P(F_s | \delta_s)$, and creates artificial spectra from the density field $\delta_s$ of a large volume $N$-body simulation by drawing flux values from $P(F_s | \delta_s)$. Here $F_s$ represents the transmitted flux field smoothed in 1-d along the line of sight by the spectral resolution of the survey being modeled, and $\delta_s$ represents the matter density field smoothed in 3-d over a scale resolved adequately in the large volume simulation. In the remainder of the paper, we drop the $s$ subscripts and use $F$ and $\delta$ to refer to the smoothed fluxes and matter density contrasts, respectively.

In this paper, as in \cite{Lochhaas2016}, we calibrate LyMAS using the Horizon simulation of \cite{Dubois2014} with no AGN feedback \citep{Peirani2017}, and we apply it to a 2048$^3$ $N$-body simulation of a (1 $h^{-1}$ Gpc)$^3$ comoving volume that is executed with GADGET2 \citep{Springel2005}. We adopt line-of-sight Gaussian smoothing of dispersion $\sigma = 0.696$ \hmpc\ comoving, appropriate to BOSS spectral resolution at $z \approx 2.5$, and 3-d Gaussian density smoothing with dispersion $\sigma = 0.5$ $h^{-1}$ Mpc. Our simulations use WMAP7 cosmological parameters \citep{Komatsu2011}, where $\Omega_m = 0.272$, $\Omega_\Lambda = 0.7284$, $\Omega_b = 0.045$, $h = 0.704$, $\sigma_8 = 0.81$, and $n_s = 0.967$. We expect that changing to Planck cosmological parameters would have a small impact on our predicted 2PCF and 3PCF but would not qualitatively change our conclusions. For further details, see \cite{Lochhaas2016}. 

The fundamental assumption of LyMAS is that any correlation between the fluxes arises only from the correlation of the underlying matter distribution. In other words, each draw of the flux value from the conditional PDFs $P(F | \delta)$ is independent, implying
\begin{equation}
P(F_1, F_2 | \delta_1, \delta_2) = P(F_1 | \delta_1) P(F_2 | \delta_2). 
\label{eqn:ansatz}
\end{equation}
\noindent 
This approximation breaks down on small scales but becomes more accurate at large separations \citep{Peirani2014}.

\cite{Peirani2014} focused on calculating the flux joint conditional PDFs \citep{Miralda-Escude1997} as a model statistic. For calculating flux correlation functions, LyMAS can be simplified. The flux 2PCF can be written generally as 
\begin{equation}
\langle \delta^F_1\delta^F_2 \rangle = \frac{\langle F_1 F_2 \rangle - \langle F \rangle^2}{\langle F \rangle^2},
\end{equation}
\noindent
with
\begin{align}
\langle F_1 F_2 \rangle 
&= \iiiint F_1 F_2 \cdot P(F_1, F_2 | \delta_1, \delta_2) dF_1dF_2 \cdot \nonumber \\
& \qquad\qquad\; P(\delta_1,\delta_2) d\delta_1 d\delta_2
\end{align}

\noindent This expression has no approximations -- we can compute $\langle F_1 F_2 \rangle$ by integrating over the full joint PDF of the matter density contrasts $\delta_1$, $\delta_2$ and over the full conditional joint PDF of the fluxes given $\delta_1$, $\delta_2$. We can now apply the LyMAS ansatz of equation (\ref{eqn:ansatz}) to write
\begin{align}
\langle F_1 F_2 \rangle 
& = \iiiint F_1 P(F_1 | \delta_1) dF_1 \cdot  F_2 P(F_2 | \delta_2) dF_2 \cdot \nonumber \\
& \qquad\qquad\; P(\delta_1,\delta_2) d\delta_1 d\delta_2 \\
& = \iint \bar{F_1}(\delta_1) \bar{F_2}(\delta_2) P(\delta_1,\delta_2) d\delta_1 d\delta_2, 
\end{align}

\noindent where the conditional mean flux is 
\begin{equation}
\bar{F}(\delta) = \int F \cdot P(F|\delta) dF.
\end{equation}

\noindent We therefore obtain the same 2PCF if we deterministically assign fluxes to $N$-body pixels using the conditional mean $\bar{F}(\delta)$ and if we draw from the full conditional $P(F | \delta)$. Averaging over pixel pairs from the simulated density field performs the integral over $P(\delta_1, \delta_2) d\delta_1 d\delta_2$ by Monte Carlo integration. A similar argument holds for the 3PCF. We have confirmed numerically that conditional mean fluxes yield the same flux correlation functions as draws from the conditional PDFs, except for the impact of random fluctuations on the latter. Using the conditional mean flux rather than draws from $P(F|\delta)$ has the advantage of producing spectra  that are coherent along the line of sight, removing the need for the ``percentile field'' mapping of \cite{Peirani2014} to create smooth mock spectra. The `full LyMAS' prescription of \cite{Peirani2014} also rescales the Fourier components of the flux field to reproduce the 1-d flux power spectrum of the hydrodynamic simulation, but we omit this step here.

\begin{figure*}
\centering
\includegraphics[width=0.85\textwidth]{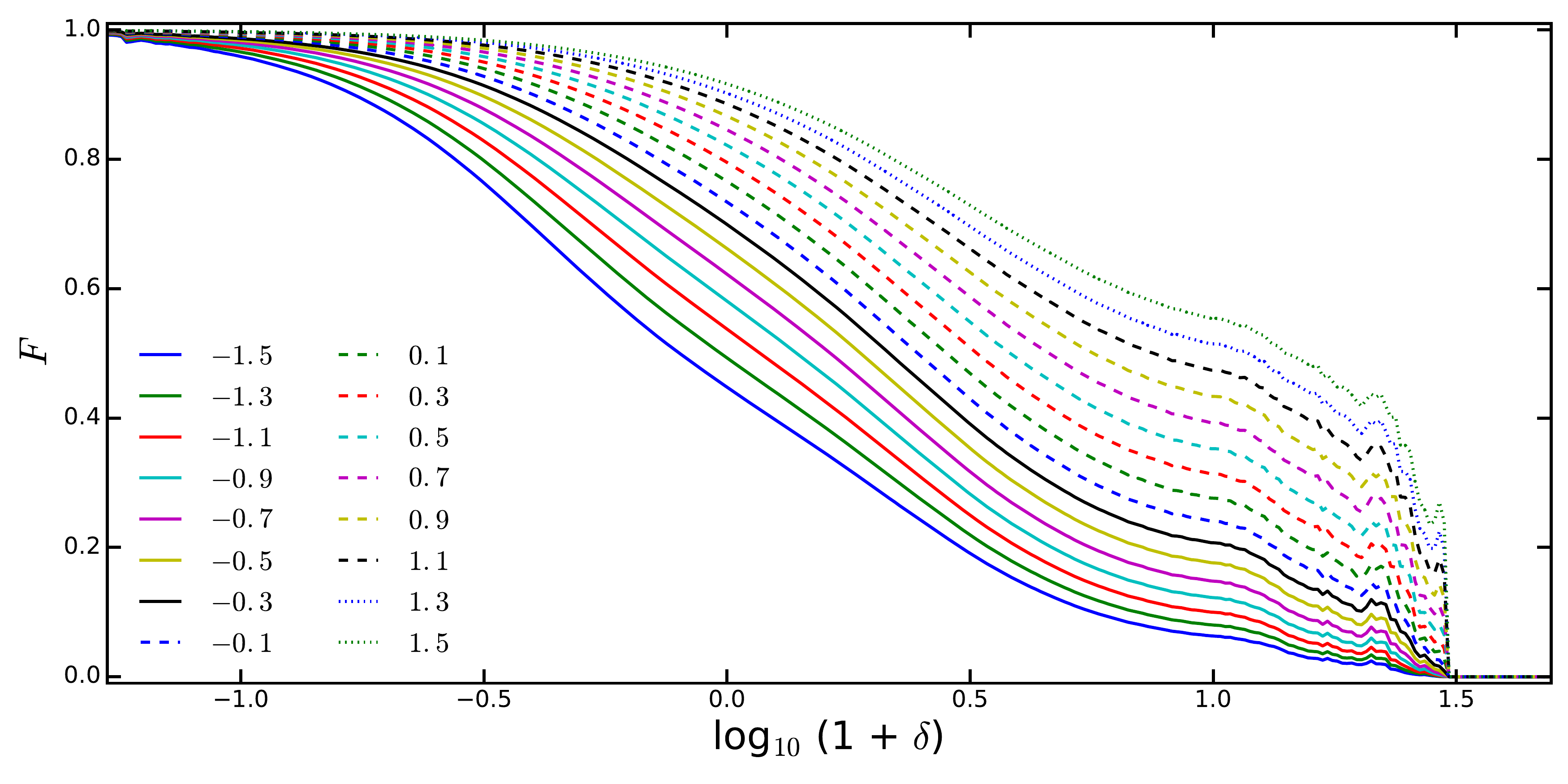}
\caption{Conditional mean flux $F$ as a function of dark matter overdensity for different UV backgrounds. The UV background is denoted by the different colored lines and ranges from ln($\Gamma/\Gamma_{0}$) = $-1.5$ to ln($\Gamma/\Gamma_{0}$) = 1.5. We only show a subset of the UV backgrounds here for brevity. The Ly$\alpha$ forest fluxes and dark matter densities are obtained from the 100 $h^{-1}$ Mpc Horizon-noAGN hydro simulation. We use a grid of DM overdensities from log$_{10}(1 + \delta) = -1.3$ to 1.7 with a step size of $\Delta(\mbox{log}_{10}(1 + \delta)) = 0.1$ to compute the conditional mean flux. As the DM grid only ranges from log$_{10}(1 + \delta)$ = $-1.275$ to 1.477, we set all fluxes to be zero for log$_{10}(1 + \delta) > 1.447$ and one for log$_{10}(1 + \delta) < -1.275$.} 
\label{lookup_table1}
\end{figure*}

We calibrate the conditional mean flux $\bar{F}(\delta)$ using the $z=2.3$ output of the Horizon-noAGN simulation \citep{Dubois2014,Peirani2017}, a (100 \hmpc)$^3$ comoving volume simulated using RAMSES \citep{Teyssier2002} in which the initially uniform grid is adaptively refined down to 1 proper kpc at all times, then sampled on a 256$^2$ grid of sight-lines with the box $z$-axis taken as the line of sight. For our uniform UVB simulation, we choose an HI photoionization rate $\Gamma_0$ that yields a mean flux $\bar{F}(z) = 0.80$ averaged over all sight-lines, in agreement with observational estimates \citep{Faucher-Giguere2008,Becker2013}. We also calibrate $\bar{F}(\delta)$ for other choices of the ionizing background intensity $\Gamma$, sampling values of ln($\Gamma_0/\Gamma$) = $-1.5$ to 1.5 with a separation of 0.1. To do so we rescale the optical depth $\tau = - \mathrm{ln }F$ of the full resolution Horizon-noAGN spectra by $\Gamma_0/\Gamma$, then apply the 0.696 \hmpc\ line-of-sight smoothing to these rescaled spectra. This method assumes that the neutral hydrogen density is inversely proportional to $\Gamma$, which is an accurate approximation for the diffuse, highly photoionized gas that produces the \lya\ forest \citep{Rauch1997,Peeples2010}. We tabulate $\bar{F}(\delta)$ at values of log$_{10}$($ 1 + \delta)$ from $-1.275$ to 1.695 in steps of 0.01, yielding a 2-d lookup table from which we can interpolate to find $\bar{F}$ at any value of $\delta$ and $\Gamma_0/\Gamma$ within the range studied. For both the Horizon-noAGN simulation and the (1 $h^{-1}$ Gpc)$^3$, 2048$^3$ $N$-body simulation, the redshift-space dark matter density field is smoothed with a 3-d Gaussian of dispersion 0.5 \hmpc\ as described by \cite{Peirani2014}. Figure \ref{lookup_table1} shows $\bar{F}(\delta)$ for a subset of our ln($\Gamma_0/\Gamma$) values. 

When applied to the dark matter density field of the calibrating hydrodynamic simulation, LyMAS reproduces the 2PCF of the full hydro spectra well but not perfectly, with the largest deviations arising for separations that are elongated along the line of sight (\citealt{Peirani2014}, Figure 20). Perturbation theory treatments of the \lya\ forest consider separate bias factors associated with the density contrast and the line of sight velocity gradient $\eta$ \citep{McDonald2003,Seljak2012}, and it may be possible to improve LyMAS by calibrating fluxes conditioned on both $\delta$ and $\eta$. We leave such an investigation to future work and for this paper note that our predicted amplitudes of the 2PCF and 3PCF could be inaccurate at the $20-30\%$ level based on the comparisons in \cite{Peirani2014} and our investigations with the Horizon simulation. Unfortunately, the hydrodynamic simulation volume is itself too small to characterize this inaccuracy with precision. 

We expect that our qualitative conclusions about the dependence of $Q$ on scale and triangle shape and the influence of UVB fluctuations on $\zeta, \xi$ and $Q$ to hold despite this quantitative uncertainty. LyMAS should be considerably more accurate than calculations based on applying the fluctuating Gunn-Peterson approximation (FGPA) to a large $N$-body simulation (e.g., \citealt{Slosar2009}), which would effectively replace the curves in Figure \ref{lookup_table1} with  linear relations between $-\mathrm{ln}F$ and log$_{10}$($1 + \delta)$. The essential problem with the FGPA for large volume simulations is that the tight relation between optical depth and matter density holds at the Jeans scale of the diffuse IGM, but does not hold between the smoothed matter density contrast and the smoothed \lya\ forest spectrum (see \citealt{Peirani2014}, Figure 4). We therefore regard LyMAS as the most promising method to make predictions for non-linear 3-d structure in the \lya\ forest at the $20-50$ \hmpc\ scales probed well by BOSS and DESI, since full hydrodynamic simulations of the requisite resolution and volume remain impractical. 

To create simulated \lya\ forest spectra with a fluctuating UVB, we first compute the quantity ln($\Gamma_0/\Gamma$) on a uniform 3-d grid of 20 \hmpc\ spacing in the 1 $h^{-1}$ Gpc simulation cube using the method described below in \S3.2, where $\Gamma_0$ represents the mean photoionization rate averaged over all points in the grid. At each pixel along each spectrum, we compute ln($\Gamma_0/\Gamma$) by linear interpolation among the surrounding grid points, then assign the value of $F$ by linear interpolation on our 2-d table of $\bar{F}(\mathrm{log}_{10}(1 + \delta), \mathrm{ln}(\Gamma_0/\Gamma))$. We apply a final multiplicative scaling of all $\Gamma$ values such that the mean flux along all spectra is again $\bar{F} = 0.8$.

\subsection{Implementing a fluctuating ionizing background}
To obtain a fluctuating radiation field, we assume quasars as our ionizing sources and place them either randomly in the box or in a random subset of massive DM halos. For the clustered quasar population, we use the DM halos identified by \cite{Lochhaas2016} using a friends-of-friends algorithm \citep{Davis1985}. We place quasars in halos with $M_{h} \geq$ $3\times10^{12}$ $M_{\odot}$, consistent with the host halo mass inferred from the clustering of BOSS quasars \citep{Ribera2013,Eftekharzadeh2015}. This mass cut selects $\sim$ 97,000 halos in the (1 $h^{-1}$ Gpc)$^3$ simulation volume. For our fiducial fluctuating UVB model, we adopt a quasar duty cycle of 10\%, i.e., we randomly select 10\% of these halos to represent active quasars at $z=2.3$. This random selection results in 9606 quasars in the box, which is a comoving volume density of $n_{q} \approx 10^{-5}$ $h^{3}$ Mpc$^{-3}$.

Comparing the clustering results for randomly placed quasars and quasars in massive halos allows us to separate the impact of shot noise and quasar clustering (see \cite{Gontcho2014} and \cite{Pontzen2014} for analytic discussion). In both cases we have simplified reality by assigning all quasar sources the same luminosity rather than drawing from a luminosity function. For randomly distributed quasars of constant luminosity $L_q$ and mean volume density $n_q$, the mean and variance of the total luminosity emitted in a volume $V$ are $V n_q L_q$ and $V n_q L_q^2$, respectively, because the variance in quasar number for a Poisson distribution is equal to the mean. For randomly distributed quasars drawn from a luminosity function $\phi(L)$, the mean and variance are $V \cdot \int_{0}^{\infty} L \phi(L) dL$ and $V \cdot \int_{0}^{\infty} L^2 \phi(L) dL$. Taking the quasar luminosity function of \cite{Kulkarni2018} at $z=2.5$, a double power-law with $\phi_{*} \approx 10^{-6}$ ($h^{-1}$ Mpc)$^3$, $\alpha \approx -4$, $\beta \approx -1.75$ (see their Figure 4), we find an rms fractional fluctuation of 0.292 $(V \phi_*)^{-1/2}$, for $V$ in comoving ($h^{-1}$ Mpc)$^3$, which is equal to that of a constant $L_q$ population of volume density $n_q = 1.17 \times 10^{-5}$ ($h^{-1}$ Mpc)$^3$. Our fiducial  case of $n_q \approx 10^{-5}$ ($h^{-1}$ Mpc)$^3$ should therefore be representative of the UVB fluctuations expected from the observed quasar population at this redshift. 

We also vary the space densities for random and clustered quasar populations by a factor of eight higher and lower to map out the dependence of the 2PCF and 3PCF on the UVB emissivity fluctuations. The contribution of galaxies to the UVB at this redshift ($z=2.3$) is uncertain, but it could potentially be non-negligible \citep{HM2012,Khaire2019a}. If galaxies make a large contribution to the UVB, then the UVB would be smoother than our $n_q = 8 \times 10^{-5}$ ($h^{-1}$ Mpc)$^3$ case, since the shot noise would be lower and the clustering bias of galaxies is weaker than that of quasars. 

The other critical parameter controlling UVB fluctuations is the mean free path $\lambda$ of ionizing photons. A smaller $\lambda$ implies that the ionizing flux at a given location comes from a smaller number of sources and is therefore subject to larger fluctuations. The mean free path is challenging to estimate observationally because absorption is dominated by systems with $\tau \sim 1$ at the Lyman limit, and these systems are relatively rare ($\sim 1$ per quasar sight-line) and their column  densities are difficult to measure because their \lya\ absorption is saturated. \cite{OMeara2013} find $\lambda \sim 570$ \hmpc\ at $z=2.44$, and \cite{Fumagalli2013} find $\lambda \sim 300$ \hmpc\ at $z=3.0$ (see \cite{Worseck2014} for a broader compilation). For our calculations, we consider $\lambda$ = 300, 100, and 50 \hmpc . Our $\lambda$ = 300 \hmpc\ is closest to (but larger than) than observational estimates near $z=2.3$, while the smaller values help illustrate behavior with stronger UVB fluctuations, which is useful for intuitive understanding and may be relevant at higher redshifts. It would be useful to have results for a still larger value of $\lambda$, but even our 1 $h^{-1}$ Gpc box is not large enough to do this. 

We assume that quasars are radiating isotropically at a constant luminosity $L$, so that the photoionization rate from quasar $i$ located a distance $d_i$ away from a point $(x, y, z)$, including periodic boundary conditions, is given by 
\begin{equation}
\Gamma_i(x, y, z) \propto L\frac{e^{-d_i/\lambda}}{4\pi d^2_i}.
\end{equation}

\noindent The value of $L$ is fixed implicitly by choosing the mean ionization rate to yield $\bar{F} = 0.8$ averaged over all sight-lines. We do not account for clustering of absorbers in the same large scale structure that hosts the quasars and the \lya\ forest, as this would require a much more complex radiative transfer calculation. In the analytic treatment of \cite{Gontcho2014} and \cite{Pontzen2014}, the impact of absorbers is roughly equivalent to modifying the quasar bias factor, so results with clustered Lyman limit absorption might be intermediate between our clustered and random quasar cases. However, a fully non-linear calculation with clustered absorption remains a goal for future work. 

\begin{figure*}
\centering
\includegraphics[width=0.9\textwidth]{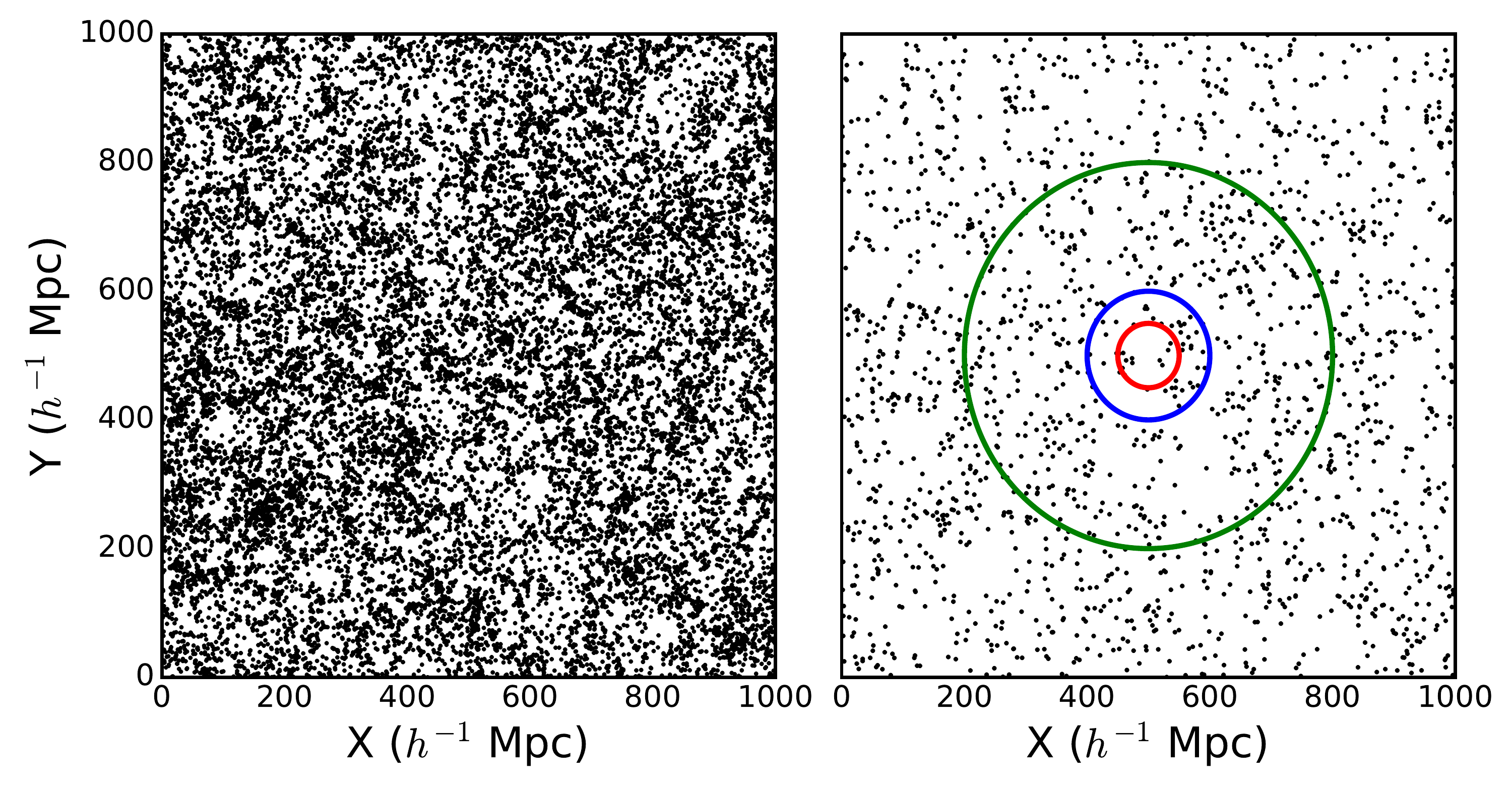}
\caption{\textit{Left}: Distribution of host halos with $M_h \geq 3\times 10^{12}$ $M_{\odot}$ in our (1 $h^{-1}$ Gpc)$^3$ box, for a slice of $\Delta Z = 200$ $h^{-1}$ Mpc. There are 96,733 DM halos that pass the mass cut in the entire box. \textit{Right}: Distribution of quasars in the same $\Delta Z$ slice after applying a 10\% duty cycle on the halos, i.e. our fiducial case with $n_q \approx 10^{-5}$ (\hmpc)$^3$. The circles denote the three mean free paths used in our UVB models, $\lambda = 300$ $h^{-1}$ Mpc (green), $\lambda = 100$ $h^{-1}$ Mpc (blue), and $\lambda = 50$ $h^{-1}$ Mpc (red). The ionizing flux at the center of the plot would be dominated by quasars within a sphere of this radius. }
\label{halo_dist}
\end{figure*}

\begin{figure*}
\centering
\includegraphics[width=1.05\textwidth]{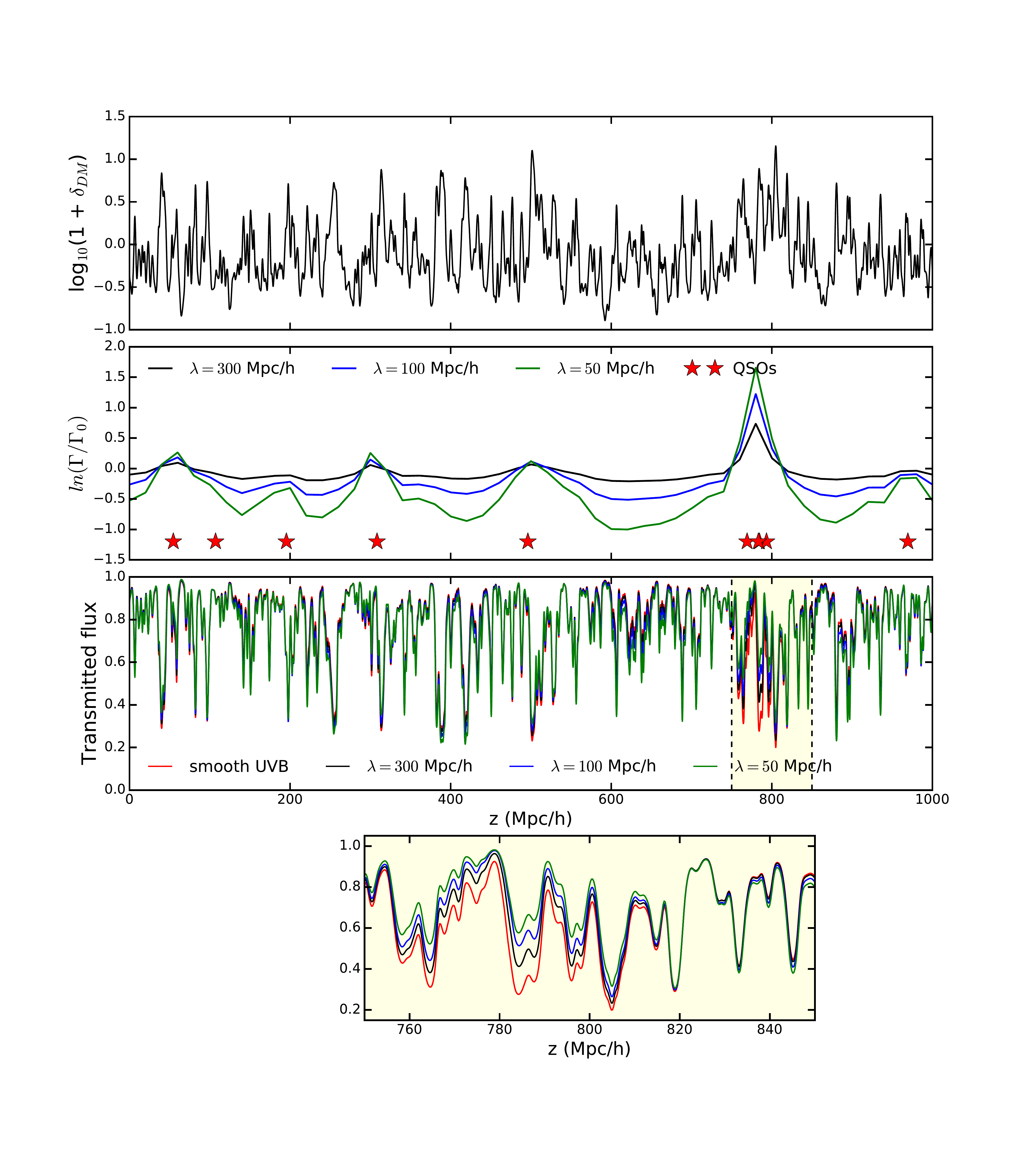}
\caption{Distribution of various physical quantities along a sight-line through our simulation with quasars in massive halos. Panels from top to bottom refer to a DM skewer smoothed to 0.5 $h^{-1}$ Mpc, relative UV intensity, transmitted flux, and a zoomed-in portion of the transmitted flux, respectively. Note that the fluxes shown here are determined by both the density and the local radiation field. The different lines refer to different mean free paths of the UV photons and the asterisks denote where the quasars are located within $\sim$ 20 $h^{-1}$ Mpc of this sight-line. While density fluctuations drive most of the structure in the forest, this structure is modulated by the UVB fluctuations. The zoom panel shows a region where higher than average $\Gamma$ reduces the absorption in the forest, an impact that is largest for the shortest mean free path.} 
\label{gamma_spectra}
\end{figure*}

Figure \ref{halo_dist} shows the distribution of quasars in massive halos for a slice in the 1 $h^{-1}$ Gpc box for our fiducial 10\% duty cycle. The UVB flux at a given location is dominated by the nearest number of quasars, $N_q \sim (4\pi/3)\lambda^3n_q \sim 1131 (n_q/10^{-5})(\lambda/300)^3$. Figure \ref{gamma_spectra} shows the combined effect of density and UVB fluctuations on the transmitted flux for a random selected sight-line through the box. As expected, the fractional flux variations ln($\Gamma/\Gamma_0$) become much larger for the shorter $\lambda$ values. Although the structure of the \lya\ forest spectrum is imprinted principally by the density fluctuations, it is modulated by the UVB fluctuations. Near $z$=800 \hmpc, a large scale overdensity is also a location of a concentration of quasars and thus a peak in the UVB intensity. The \lya\ forest absorption is therefore reduced relative to the uniform background case (see zoom panel), more so for the shortest $\lambda$. With clustered ionizing background sources, density and UVB fluctuations tend to have opposite impact on the \lya\ forest absorption. However, even for $\lambda = 50$ \hmpc, the scale of UVB fluctuations is much larger than that of the density fluctuations that produce order unity \lya\ flux variations. 

\begin{figure}
\centering
\includegraphics[width=0.47\textwidth]{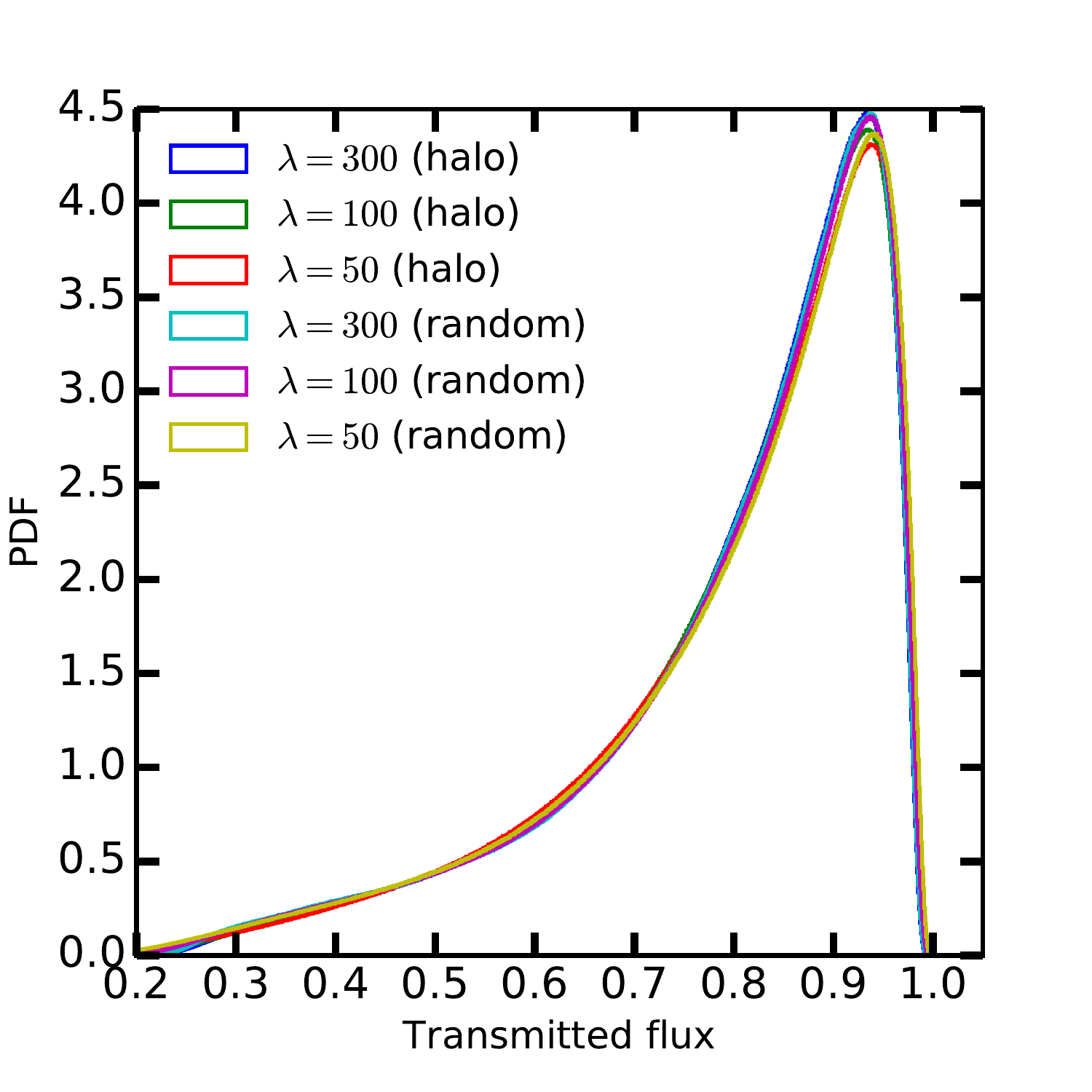}
\caption{Distribution of forest fluxes in our (1 $h^{-1}$ Gpc)$^{3}$ box for different choices of $\lambda$ for quasars found in massive halos and randomly-distributed quasars. The forest fluxes have been rescaled to the observed mean flux of $\bar{F} = 0.8$. The flux PDF is insensitive to the choices of $\lambda$.}
\label{flux_pdf}
\end{figure}

Figure \ref{flux_pdf} shows the PDF of transmitted flux from all sight-lines through the box. This PDF is remarkably insensitive to the presence of UVB fluctuations. However, we will show that these fluctuations have a significant impact on the flux 2PCF and 3PCF. 

\subsection{Calculating the Ly$\alpha$ forest clustering}
There are 65,536 sight-lines in our 1 $h^{-3}$ Gpc$^3$ $N$-body box, with a minimum sight-line separation of $ds$ = 3.91 $h^{-1}$ Mpc and each spectrum consisting of 4096 pixels. We currently only correlate sight-lines and pixels at the same redshifts (or $z$ positions, i.e. the planes of the triplets are perpendicular to the line of sight), with sight-line separations up to a maximum of 60 $h^{-1}$ Mpc. 

\begin{figure}
\centering
\includegraphics[width=0.47\textwidth]{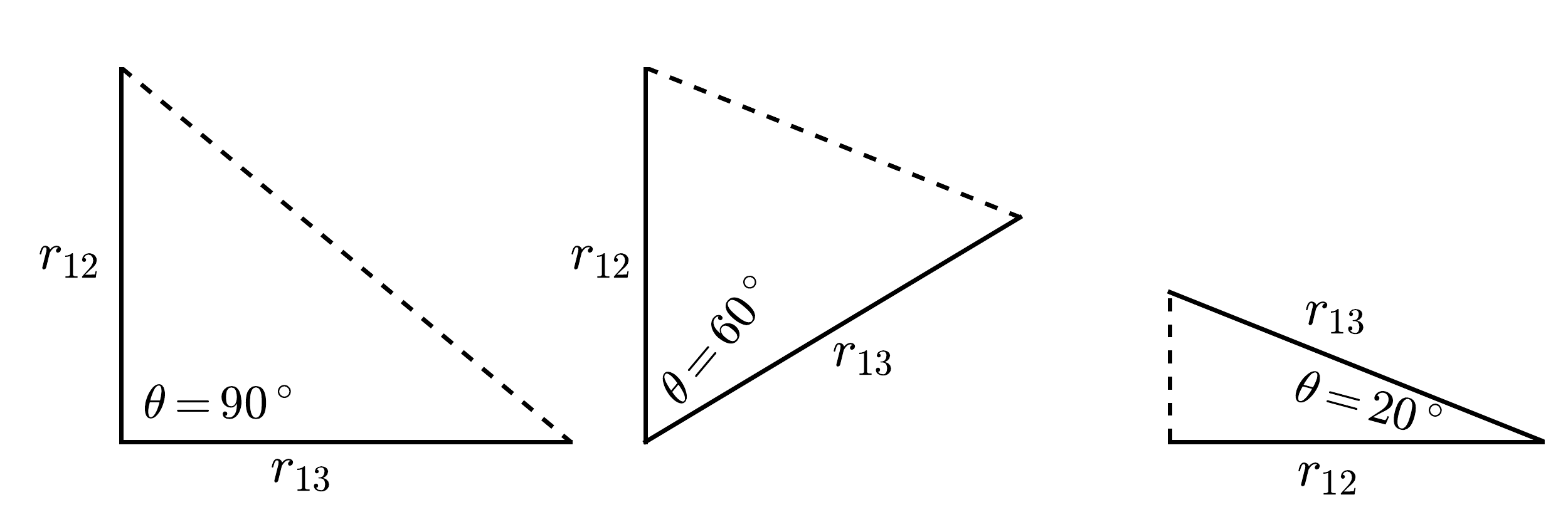}
\caption{The three triplet configurations investigated in this paper. We select triangles with $r_{13}/r_{12}$ = 1 and with opening angle $\theta = 90^\circ$ (left), $\theta = 60^\circ$ (middle), and $\theta = 20^\circ$ (right), where $\theta$ is the angle between $r_{12}$ and $r_{13}$. We allow for margins in $\theta$ of $\pm$ 5$^\circ$ and in $r_{13}/r_{12}$ of $\pm$ 0.2.} 
\label{eg_triplets}
\end{figure}

\begin{figure}
\centering
\includegraphics[width=0.45\textwidth]{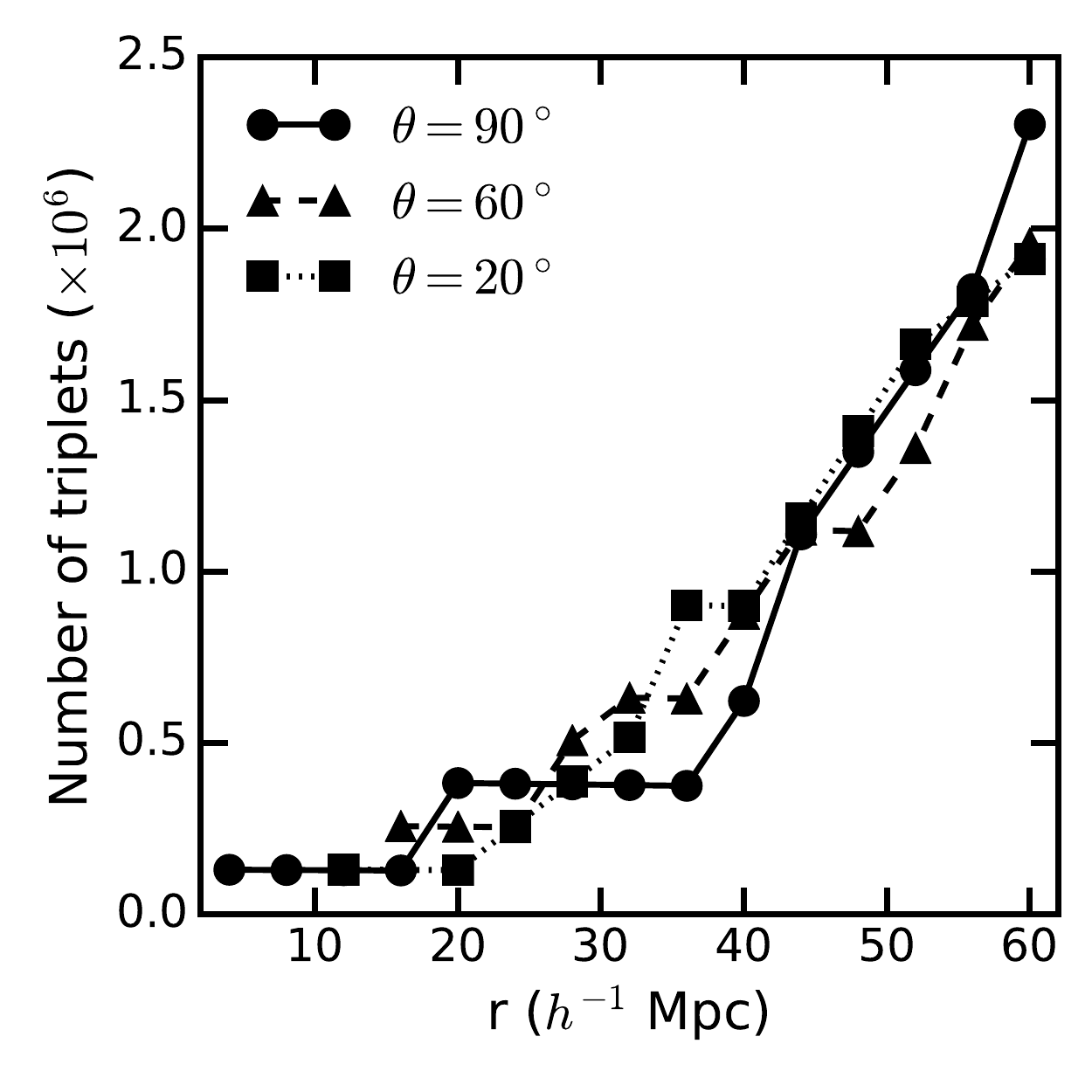}
\caption{The total number of sight-line triplets sampled at each separation. The different lines indicate different triplet configurations. The number of available sight-lines increases with separation. } 
\label{ntriplets}
\end{figure}

We select triplets with roughly equal side lengths, $r \sim r_{12} \sim r_{13}$ and for three different triangle opening angles $\theta$ = 90$^{\circ}$, 60$^{\circ}$, and 20$^{\circ}$, each with an angle margin of $\pm$ 5$^{\circ}$. Recall that $\theta$ is defined as the angle between the vectors $r_{12}$ and $r_{13}$. We choose $r_{13}$ separations spanning from 0.8$r_{12}$ to 1.2$r_{12}$, where $r_{12} = ds, 2ds, 3ds ... 15ds = 3.91, 7.82, 11.73 ... 58.65$ \hmpc\ . For each $r_{12}$, we iteratively use every sight-line in the box as a primary sight-line, then randomly select one of the four possible second sight-line located $r_{12}$ away, and finally select all possible third sight-lines to complete the triplet within the $r_{13}$ and angle ranges. Since sight-lines are repeated, the error bars of the 2- and 3-point correlation functions at various scales are correlated. Figure \ref{eg_triplets} shows an example of a triplet configuration for each $\theta$, and Figure \ref{ntriplets} shows the total number of triplets in our box as a function of separation. 

We calculate the 2PCF, 3PCF, and $Q$ of the triplets according to Equations (1), (3) and (4), in bins of 4 $h^{-1}$ Mpc. The final $Q$ values are obtained using all sight-lines in the entire box. To estimate the error bars, we divide the triplets into nine subvolumes and calculate the 2PCF and 3PCF using all sight-lines in each subvolume. The subvolumes are divided in $X$ and $Y$, but not in $Z$, so each subvolume is essentially a long narrow rectangular prism. The reduced 3PCF for a subvolume $i$, $Q_i$, is obtained accordingly using the respective correlation functions.

\begin{align}
Q_i(r) &= \frac{\zeta_{123,i}}{(\xi_{12,i})(\xi_{13,i}) + (\xi_{12,i})(\xi_{23,i}) + (\xi_{13,i})(\xi_{23,i})}\\
\zeta_{123,i} &= \frac{1}{N_i(r_b)} \Sigma_{j=1}^{N_i(r_b)} \delta^F_1(x_1)\delta^F_2(x_2)\delta^F_3(x_3)\\
\xi_{12,i} &= \frac{1}{N_i(r_b)} \Sigma_{j=1}^{N_i(r_b)} \delta^F_1(x_1)\delta^F_2(x_2) \\
\xi_{13,i} &= \frac{1}{N_i(r_b)} \Sigma_{j=1}^{N_i(r_b)} \delta^F_1(x_1)\delta^F_3(x_3) \\
\xi_{23,i} &= \frac{1}{N_i(r_b)} \Sigma_{j=1}^{N_i(r_b)} \delta^F_2(x_2)\delta^F_3(x_3) 
\end{align}

\noindent
where $N_i(r_b)$ is the number of triplets in each subvolume $i$ in bin $r_b$. We estimate our error bars as the standard deviation in the correlation functions among the subvolumes divided by the square root of the number of subvolumes. They therefore represent our estimate of the uncertainty in the theoretical prediction from the full (1 $h^{-1}$ Gpc)$^3$ simulation volume. We discuss the source of statistical uncertainty in our predictions below, especially in Appendix A, concluding that it is dominated by cosmic variance of large scale structure within our survey volume. 

\section{Results}
\subsection{Reduced 3PCF for a uniform ionizing background}
\begin{figure}
\centering
\includegraphics[width=0.43\textwidth]{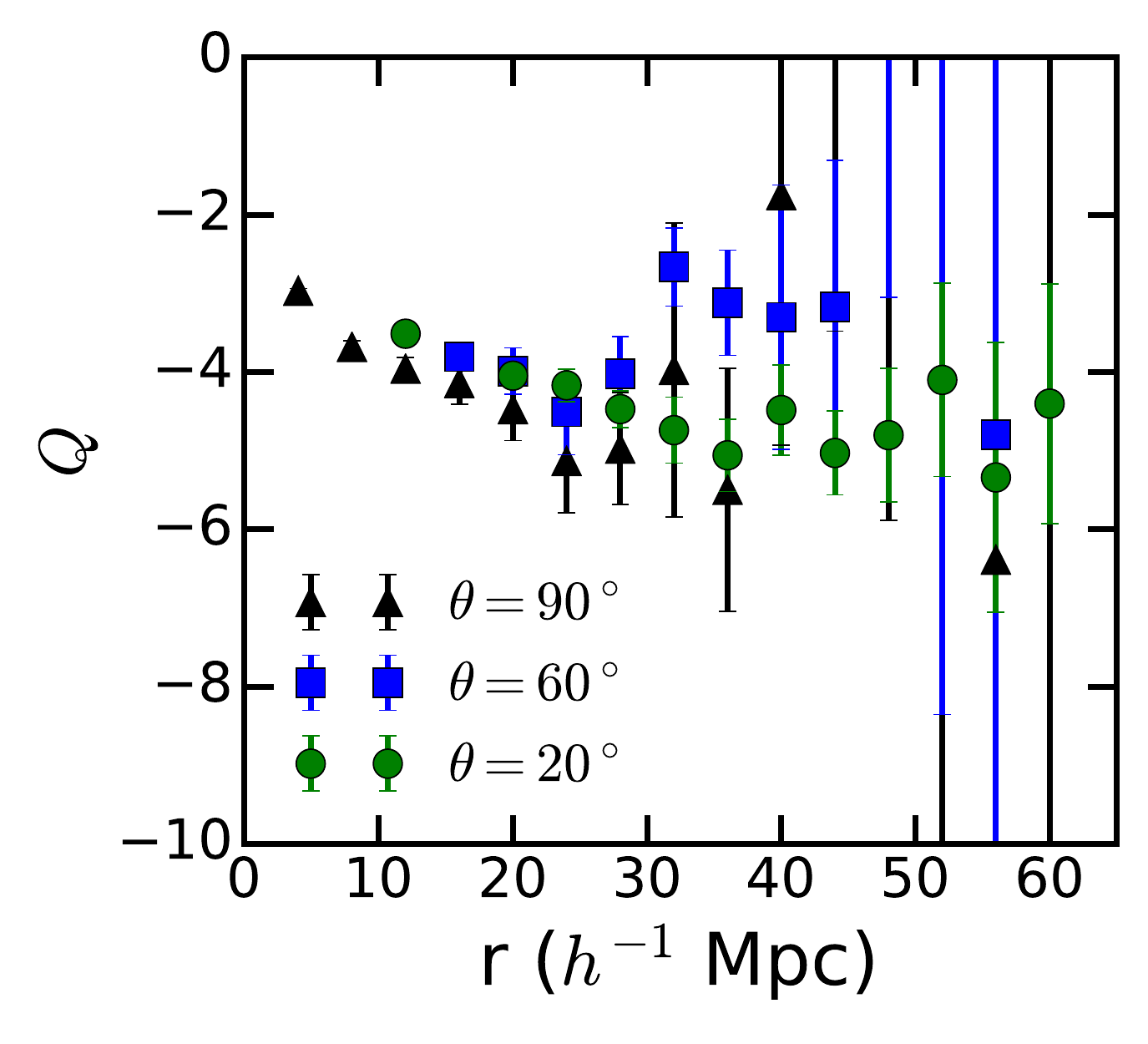}
\caption{The reduced 3PCF $Q$ as a function of separation $r$ in a smooth ionizing background. The different colored points refer to the different triangle shapes. The $Q$ of the \lya\ forest does not show a clear trend with shape and is approximately constant at $\sim -4$. The error bars are computed from dividing our box into subvolumes as described in \S 3.3. } 
\label{Qsmooth}
\end{figure}

We show the reduced 3PCF in a uniform ionizing background as a function of $r$ for all three triangle shapes we investigated in Figure \ref{Qsmooth}. The $Q$ value has little dependence on the triangle shape and remains approximately constant at $\sim -4$ despite the 3PCF changing by more than two orders of magnitude (see Figure \ref{halo-3pcf}). Compared to galaxies, which have a positive and small $Q_m$ ($\approx 1.3$) \citep{Peebles1975,Groth1977}, the value of $Q$ for the \lya\ forest is negative and large. The negative value of $Q$ (and the 3PCF) arises because the forest is in absorption, so that high density produces low flux. The large amplitude of $Q$ reflects the low bias factor of the forest. With a local bias model of the forest at second order, 
\begin{align}
Q = \frac{Q_m}{b} + \frac{b_2}{b^2}
\end{align}

\noindent we assume the forest flux fluctuation is related to the DM overdensity by 

\begin{align}
\delta^F = b\delta + \frac{b_2}{2}\delta^2 - \frac{1}{2}b_2\langle \delta^2 \rangle 
\end{align}

\noindent \citep{Fry_Gaztanaga1993,Fry1994,Juszkiewicz1995}. We get $Q = -7.6$ for $b_2 = 0$ when adopting $Q_m = 1.3$ and $b = -0.17$ \citep{Slosar2011}; reproducing our simulation value of $Q \sim 4$ requires $b_2 \approx 0.1$. Thus small values of $b$ and $b_2$ for the forest naturally give rise to a large value of $Q$.

\subsection{Impact of ionizing radiation fluctuations and source clustering}
\begin{figure*}
\centering
\includegraphics[width=0.85\textwidth]{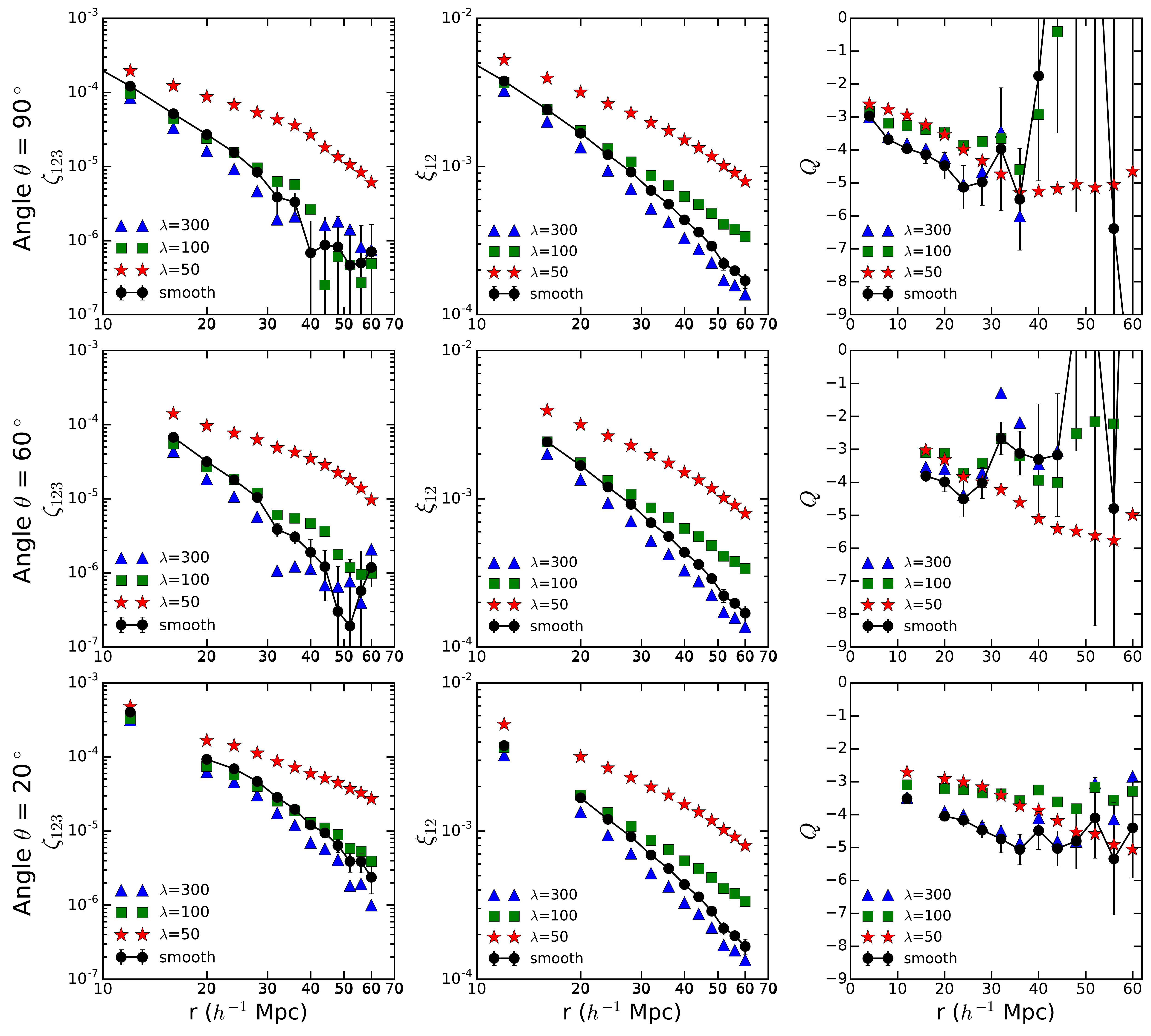}
\caption{The clustering of the Ly$\alpha$ forest for quasars in massive halos ($M_h \geq 3 \times 10^{12}$ $M_{\odot}$) with different radiation mean free path $\lambda$. The cut on the halo masses results in a volume density of $n_{q} \sim 10^{-5}$ $h^{3}$ Mpc$^{-3}$. A non-fluctuating smooth background is shown as the connected black points. The triplet configurations are shown as the different rows. As the 2PCF trends of the three side lengths are similar, we only show the 2PCF of one side length. We also only show the error bars for the smooth background to avoid crowding, but they are similar for the fluctuating backgrounds. Our measurements at the different scales are correlated because sight-lines are repeated for triplets at different separations (see \S3.2). For the $\theta = 20^\circ$ triangles with $r_{23} < r_{12} \sim r_{13}$, $\zeta_{123}$ is higher, but $Q$ is similar in amplitude to other triangle shapes because in the hierarchical normalization $\xi_{23}$ is larger than $\xi_{12}$ and $\xi_{13}$.}
\label{halo-3pcf}
\end{figure*}

\begin{figure*}
\centering
\includegraphics[width=0.85\textwidth]{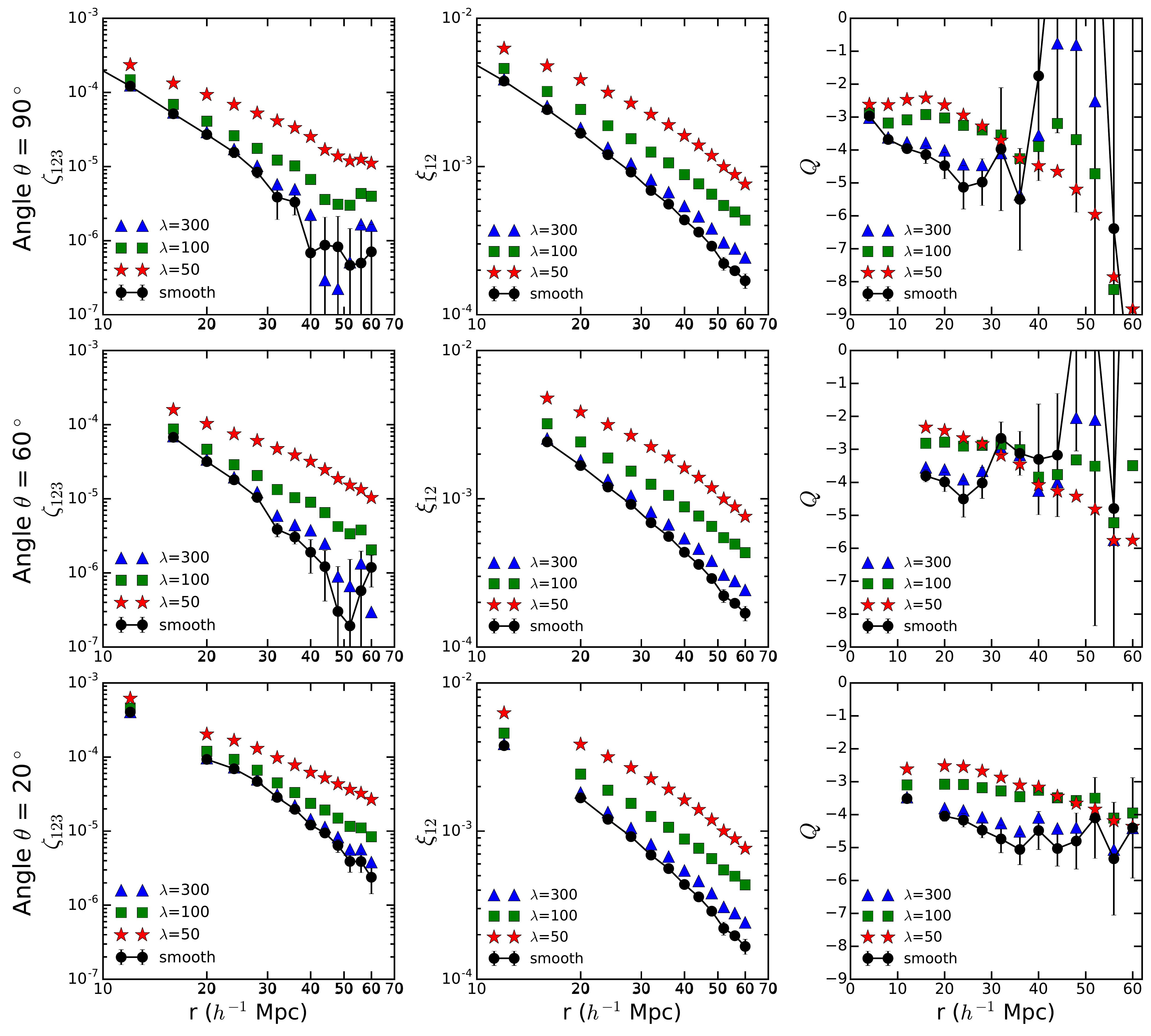}
\caption{Same as Figure~\ref{halo-3pcf}, but for a random quasar distribution with the same volume density ($n_{q} = 10^{-5}$ $h^{3}$ Mpc$^{-3}$). }
\label{random-3pcf}
\end{figure*}

We compare the 2PCF and 3PCF of the Ly$\alpha$ forest in different fluctuating UV backgrounds and with different triplet configurations of sight-lines. Figure \ref{halo-3pcf} and \ref{random-3pcf} show the results for quasars found in massive halos and for randomly-distributed quasars, respectively. UVB fluctuation changes the clustering at all scales and produces increased signal as $\lambda$ gets smaller.  Although the flux PDF remains unchanged with UVB fluctuations (Figure \ref{flux_pdf}), the 2PCF and 3PCF are clearly changed.

For a clustered quasar source population with $\lambda = 300$, the 2PCF of the forest is moderately suppressed by a factor of $\sim 1.2-2$, whereas the 2PCF from unclustered quasar sources is similar to that for a uniform background at small scales but slightly enhanced at large scales. In the unclustered source cases, flux variations are a source of additional large scale structure in the forest. For halo-based quasars on the other hand, high density regions also have higher UVB, and the cancellation suppresses clustering overall. 

For $\lambda = 100$, the UVB fluctuations are larger in amplitude, and for unclustered quasar sources the flux 2PCF and 3PCF are enhanced significantly at all scales. With clustered sources there is again partial cancellation, but at large scales the 2PCF is now enhanced relative to the uniform UVB instead of suppressed. For $\lambda = 50$, the flux 2PCF and 3PCF are dramatically enhanced at all scales; in this case the large scale clustering of the forest is dominated by UVB variations rather than gas density fluctuations. One might hope that the transition to a different clustering origin would lead to a sharp departure from the hierarchical relation of the 3PCF and 2PCF, but the $\lambda = 50$ and $\lambda = 100$ cases still have approximately constant $Q$ out to $r = 30$ \hmpc , with a moderate reduction from $|Q| \approx -4.5$ from the smooth UVB case to $|Q| \approx -3$. 

There could be significant changes in $Q$ at larger scales, but the error bars from our finite simulation volume are too large to tell. In all cases the triangle shape has only moderate impact on $Q$, though for $\theta = 20^\circ$ the error bars at large $r$ are reduced because $r_{23}$ remains relatively small, so $\zeta_{123}$ and $\xi_{23}$ are larger and better measured. For this triangle shape we see only moderate scale dependence of $Q$ out to $r = 60$ \hmpc , and the value of $|Q|$ is lower for the $\lambda = 50$ and $\lambda = 100$ cases relative to the $\lambda = 300$ and uniform cases. 

\begin{figure*}
\centering
\includegraphics[width=0.85\textwidth]{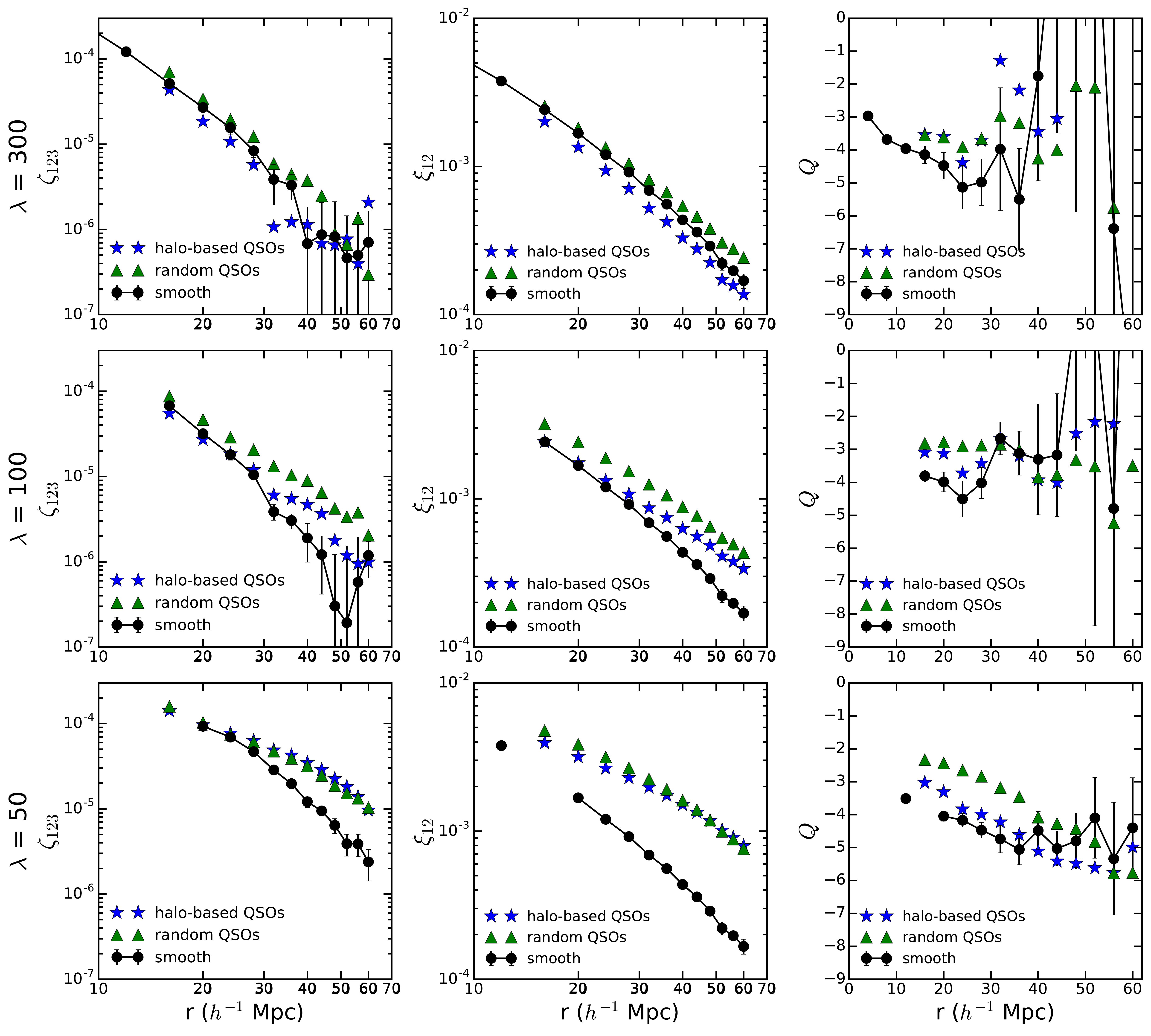}
\caption{Comparison of the Ly$\alpha$ forest clustering between quasars in massive halos and randomly-distributed quasars, both with $n_{q} \sim 10^{-5}$ $h^{3}$ Mpc$^{-3}$, for triplet configuration with $\theta = 60^\circ$. The different rows refer to the different mean free path $\lambda$. Halo-based sources typically lead to a less clustered \lya\ forest due to cancellation between increased gas density and increased local UVB intensity.}
\label{halo_vs_random}
\end{figure*}

Figure \ref{halo_vs_random} compares the impact of clustered vs. unclustered radiation sources more directly, for triplets with $\theta = 60^\circ$. Quasars in massive halos tend to produce weaker forest clustering than randomly-placed quasars for all $\lambda$ values. Because hierarchical behavior is a ``special'' consequence of gravitational instability of Gaussian initial conditions, we anticipated that we might see sharp scale-dependence of $Q$ associated with the scale of $\lambda$. However, this is not evident within our errors. Larger simulation boxes are needed to further test this conjecture. As discussed further in Appendix A, our results are stable against different random realizations of quasar distribution. Rather than being limited by the number of sight-lines or triplets in our box, our measurements are limited by the variance due to large scale structures. We therefore need larger simulation volumes rather than more complete sampling of this simulation. 

\subsection{Impact of shot noise}
Shot noise from the rarity of the ionizing sources can affect and complicate interpretation of \lya\ forest clustering measurements. We investigate the impact of shot noise by changing the number of quasars in the different fluctuating backgrounds by a factor of eight from the fiducial $N_q = 10,000$. We again assume either a halo-based or random quasar distribution. For halo-based quasars, we use the same set of DM halos and the same mass cut of $M_{h} \geq$ $3\times10^{12}$ for the quasar hosts as before, but vary the quasar duty cycle to 80\% and 1.25\%. For random quasar distributions, we specify the desired numbers of quasars exactly (either higher or lower by a factor of eight from the fiducial) and randomly assign their $(X, Y, Z)$ positions in the box. We follow the same steps listed in \S3.1 to generate the resultant \lya\ forest fluxes using the new UVB grids, making sure to renormalize the new fluxes to the observed mean flux of 0.8. The resultant flux histograms (1-point PDFs) are nearly unchanged for different quasar densities, similar to Figure \ref{flux_pdf}.

\begin{figure*}
\centering
\includegraphics[width=0.95\textwidth]{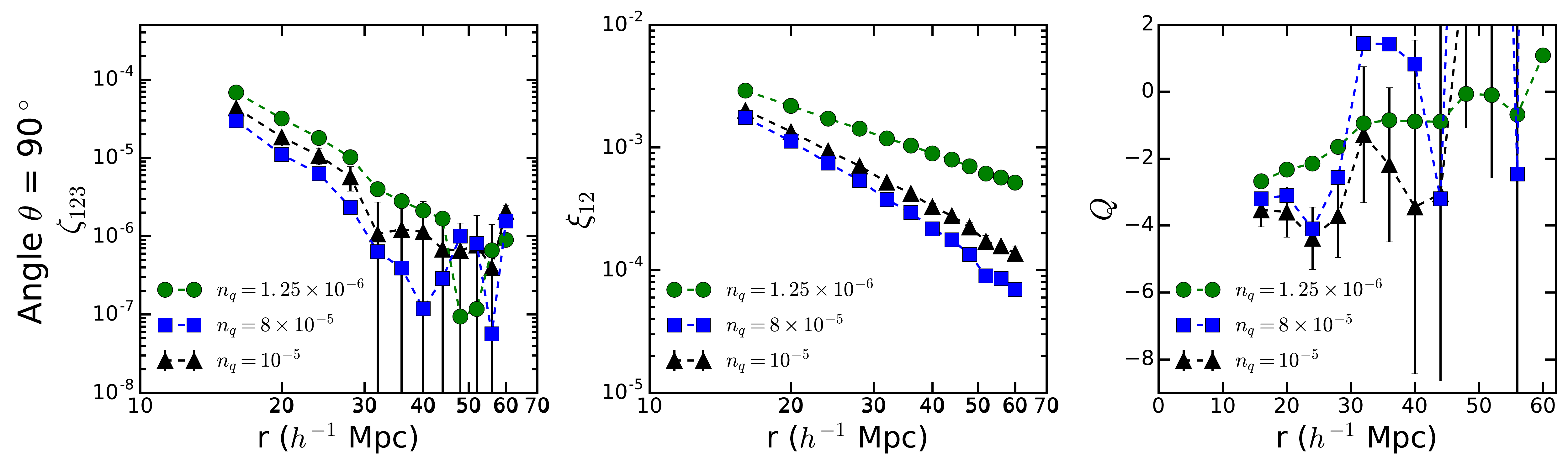}
\caption{The clustering of the \lya\ forest from clustered quasars with a fixed $\lambda = 300$ \hmpc\ and varying volume density $n_q$. The figures here are for an equilateral triplet configuration, $\theta = 60^\circ$ (other triangle shapes show similar results). The fiducial $n_q$ is given by the black points. Shot noise results in the largest clustering signal for the lowest $n_q$ while being most suppressed in the largest $n_q$. }
\label{lamb300_3nq}
\end{figure*}

Figure \ref{lamb300_3nq} shows the 2PCF, 3PCF, and $Q$ measurements for clustered quasar populations at the three volume densities for an equilateral triplet configuration in a $\lambda = 300$ \hmpc\ ionizing background. Shot noise adds broadband power on all scales, giving rise to the largest clustering signal for the lowest $n_q$ (green points) while being suppressed in the largest $n_q$ (blue points). Values of $Q$ are not sensitive to $n_q$, at least relative to our error bars. 

\begin{figure*}
\centering
\includegraphics[width=0.95\textwidth]{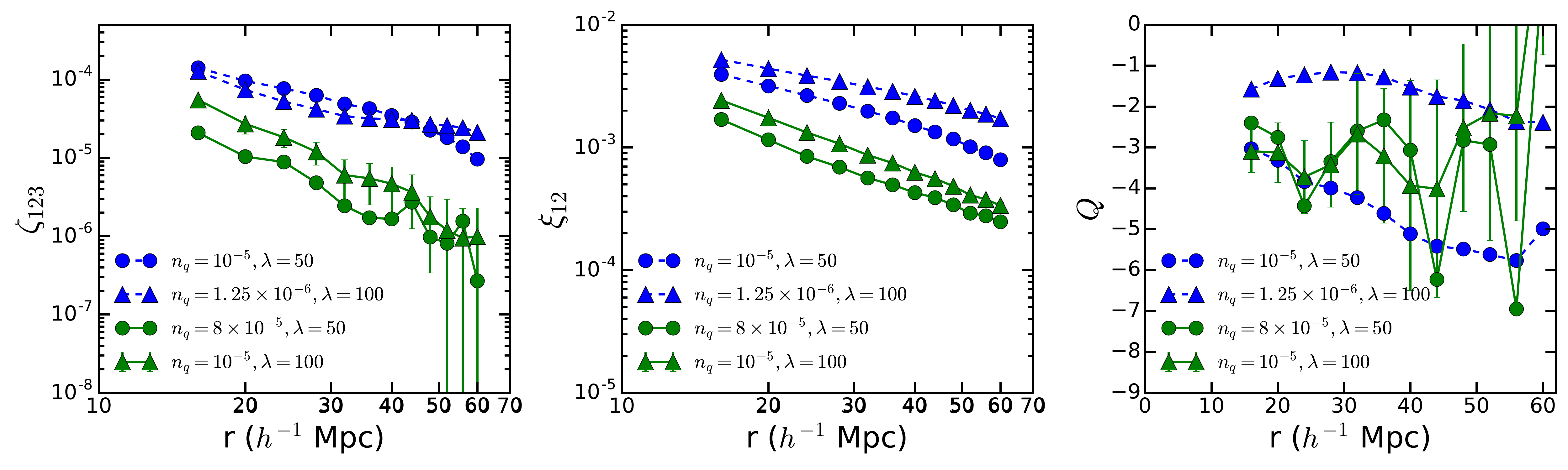}
\caption{The Ly$\alpha$ forest clustering for halo-based quasars with different combinations of volume density $n_q$ and $\lambda$. Lines with the same color have the same number of quasar $N_q$ within a volume of $\lambda^3$, $N_q (\leq \lambda^3)$, while lines with the same point shape have the same $\lambda$. The blue lines have $N_q (\leq \lambda^3) \sim 5$ while the green lines have $N_q (\leq \lambda^3) \sim 42$. This figure is for triplets with $\theta = 60^\circ$. Other triplet configurations show similar behavior. The blue lines with a lower $N_q (\leq \lambda^3)$ have a higher 2- and 3-point clustering than the green lines, as shot noise adds broadband power at all scales. Although the value of $N_q$ sets the normalization for the different combinations of $n_q$ and $\lambda$, it does not fully determine the forest clustering. }
\label{matched_all}
\end{figure*}

Figure \ref{matched_all} shows the $\lambda = 50$ and $\lambda = 100$ cases with different $n_q$. The clustering signals increase for shorter $\lambda$ or lower $n_q$ as expected. Lines with the same color show $(n_q, \lambda)$ combinations chosen to have the same $N_q = 4/3 \pi \lambda^3 n_q$; blue lines have $N_q \sim 5$ and green lines have $N_q \sim 42$. The value of $N_q$ clearly separates these combinations, though it does not fully determine the forest clustering. The case of low quasar density $n_q = 1.25 \times 10^{-8}$ and $\lambda = 100$ is the one combination that shows a very different value of $Q$ at $\sim -1.5$. One can see a suggestion of reduced $|Q|$ for low $n_q$ and $\lambda = 300$ in Figure \ref{matched_all}. Randomly placed quasars follow the same behavior and trend, except with larger correlation function values and smaller $|Q|$ amplitudes than clustered quasars, reflecting what we see in Figure \ref{halo_vs_random}.

\subsection{Detectability of the 3PCF}
\cite{Slosar2011} made the first detection of the 3-d 2PCF redshift-space distortion in the \lya\ forest. With subsequent data from BOSS, the 2PCF has been measured with increasing precision to constrain the BAO peak \citep{Busca2013,Slosar2013,Delubac2015,Bautista2017,Bourboux2017,Blomqvist2019,Agathe2019}. To our knowledge, no 3PCF measurement of the forest has been made. We attempted to measure the 3PCF using recent data from the CLAMATO 3D \lya\ forest tomography survey \citep{Lee2018}. Although we obtained a clear 2PCF signal, the 3PCF measurement is consistent with noise. We can therefore ask whether the 3PCF should be detectable in current and future surveys, e.g., in BOSS, eBOSS, and DESI. 

To answer this question, we estimate the expected signal-to-noise ratio (SNR) of a 3PCF detection using specifications that approximate these current spectroscopic surveys. The volume probed by these surveys is much larger than our 1 ($h^{-1}$ Gpc)$^3$ box. For example, the BOSS survey volume for the redshift range $z = 2.1 - 2.6$ over 10$^4$ deg$^2$ corresponds to $\sim 21$ $h^{-3}$ Gpc$^3$. However, the BOSS sampling density is far lower; given 114,600 quasars between $2.1 < z < 2.6$ distributed over an area of 9376 deg$^{-2}$ (BOSS DR12, \citealp{Alam2015}), its sampling density is $\sim$ 12 quasars deg$^{-2}$, corresponding to a comoving surface density of $\Sigma = 2.6 \times 10^{-3}$ $h^2$ Mpc$^{-2}$ at $z=2.3$. Our previous analyses use 256$^2$ sight-lines through our box, which gives a comoving surface density of $\Sigma = 6.55 \times 10^{-2}$ $h^2$ Mpc$^{-2}$. 

For our SNR estimate, we assume an observed surface density of 10 quasars deg$^{-2}$ at $z = 2.3$, or $\Sigma = 2 \times 10^{-3}$ $h^2$ Mpc$^{-2}$, roughly comparable to that of BOSS. This translates to 2000 sight-lines for our 1 $h^{-1}$ Gpc box. A typical Ly$\alpha$ forest region spanning the range from the quasar's \lya\ to Ly$\beta$ emission lines is $\sim 300$ \hmpc\ long, three times shorter than our simulation sight-lines. As we analyzed the entire sight-lines, this means we effectively have 6000 sight-lines or a 3$\times$ larger effective volume. We estimate the noise for BOSS, eBOSS, and DESI by scaling the noise from our box by $1/N^{1/2}_{\mathrm{sightline}}$. This scaling is appropriate because errors in widely separated regions (i.e., larger than our simulation box) should be independent.

After selecting 2000 random sight-lines through our box, we consider loosely equilaterial triangle configurations with a fractional width of $w = \Delta r/r$ = 0.2 at three separations of $r = 10, 20$, and 40 $h^{-1}$ Mpc. For instance, for $r$ = 10 $h^{-1}$ Mpc, we consider sight-lines that are between 8 $h^{-1}$ Mpc and 12 $h^{-1}$ Mpc away from the primary sight-line. The same fractional width is also applied when we correlate pixels from the sight-line triplets, such that we are not limited to strictly face-on pixel triplets. In other words, we correlate pixels in sight-line $i$ with pixels that are located between $z_i - wr$ and $z_i + wr$ in the other two sight-lines. 

We ran 50 realizations in which we chose 2000 random sight-lines from our uniform UVB box and measure the 3PCF using all sight-line and pixel triples that satisfy the shape criterion mentioned above. These 50 realizations have a mean of 15, 271, and 333 sight-line triplets at $r = 10, 20$, and 40 $h^{-1}$ Mpc, respectively. We take the dispersion among these 50 realizations to represent the rms error of $\zeta_{123}(r)$ expected for 6000 forest spectra with a surface density of 10 deg$^{-2}$ at $z = 2.3$. We compute the SNR as the ratio of the mean $\zeta_{123}$ to this dispersion. This calculation implicitly assumes that the statistical errors at this quasar surface density are dominated by sparse sampling of the available structure and not by cosmic variance of the structure itself. We believe this assumption is justified, but we have not demonstrated it. 

Another source of noise in observational data is photon noise in the spectra. The photon noise per pixel can be reduced by smoothing the spectra, though this also reduces the number of independent pixel triplets. To help assess this issue, we compute the SNR for our full resolution spectra, for spectra that are boxcar-smoothed over 4 or 16 pixels (each roughly comparable to a BOSS pixel), and for spectra binned over 4 or 16 pixels, which therefore have a factor of 4 or 16 smaller pixel count. To isolate the effect of smoothing from that of number of pixel triplets, we also calculate the SNR after simply choosing every 4th or 16th pixel from that full resolution spectra. 

\begin{figure*}
\centering
\includegraphics[width=0.95\textwidth]{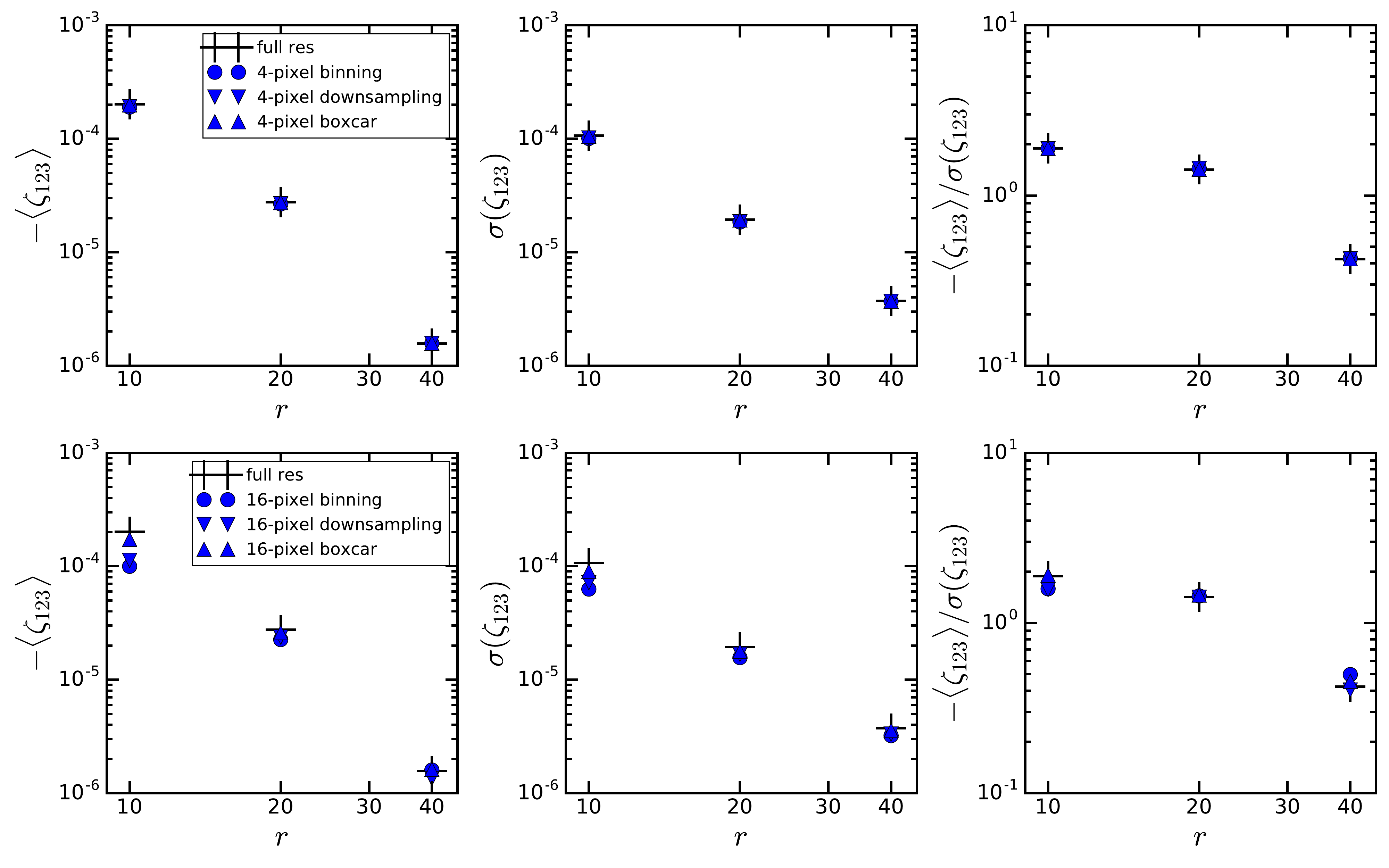}
\caption{Expected signal (left), noise (middle), and SNR (right) of a 3PCF measurement in the \lya\ forest for 50 realizations of 2000 random sight-lines through our 1 Gpc $h^{-1}$ box. The signal is the mean 3PCF of all the realizations and the noise is the dispersion of the 3PCF values among the realizations. We smooth the spectra by four pixels in the top row and sixteen pixels in the bottom row. For $n$-pixel binning, we average all pixels in a bin containing $n$ pixels; for $n$-pixel downsampling, we consider only every $n$-th pixel and disregard all pixels in between; for $n$-pixel boxcar, we run a 1D boxcar kernel across every spectrum. Binning and downsampling reduce the spectrum length by 1/$n$ while boxcar smoothing retains the original number of pixels in each spectrum. The simulation box length is roughly three times the length of the \lya\ forest in a typical high-$z$ quasar spectrum. The expected SNR for a sample of $N$ \lya\ forest spectra at this surface density should therefore be multiplied by $\sim (N/6000)^{1/2}$. }
\label{snr}
\end{figure*}

Figure \ref{snr} shows our results. For 6000 sight-lines we expect a SNR of $\sim 2$ at $r = 10$ \hmpc , $\sim 1.5$ at $r = 20$ \hmpc , and $\sim 0.4$ at $r = 40$ \hmpc . For the 114,600 sight-lines within the redshift range $z = 2.1 - 2.6$ found in BOSS DR 12 \citep{Alam2015}, we predict a SNR higher by $(114,600/6000)^{1/2} \approx 4.4$, so roughly 8.7, 6.6, and 1.75 at these separations. The eBOSS survey has a comparable surface density of 13.8 quasars deg$^{-2}$ within the redshift range $z=2.1 - 2.6$. For the 129,975 sight-lines within the above redshift range found in eBOSS DR 14 quasar catalog \citep{Paris2018}, we predict a very similar SNR of roughly 9.3, 7, and 1.9 for these separations. Remarkably, smoothing, binning, or sampling with a scale of 4 or 16 pixels has essentially no impact on the SNR. This is encouraging as it implies that one could reduce photon noise by smoothing up to 16 pixels affecting the SNR. One can therefore either bin or average a larger number of (individually noisier) pixel triplets to reduce photon noise in the observed spectra. Binning has the additional attraction of reducing the computational demands of the 3-point measurement without loss of sensitivity. 

We next consider the effect of changing the quasar surface density on the SNR by repeating our simulations for 4000 and 8000 sight-lines, corresponding to increasing the quasar surface density by two (giving 20 quasars deg$^{-2}$) and four (giving 40 quasars deg$^{-2}$). As a comparison, the DESI survey will have a surface density of 50 quasars deg$^{-2}$ at $z > 2.1$. For the redshift range $z=2.1 - 2.6$, its surface density is estimated to be 30 quasars deg$^{-2}$ (based on Figure 3.17 of \citealp{DESI2016}), resulting in 420,000 total sight-lines.

\begin{figure*}
\centering
\includegraphics[width=0.95\textwidth]{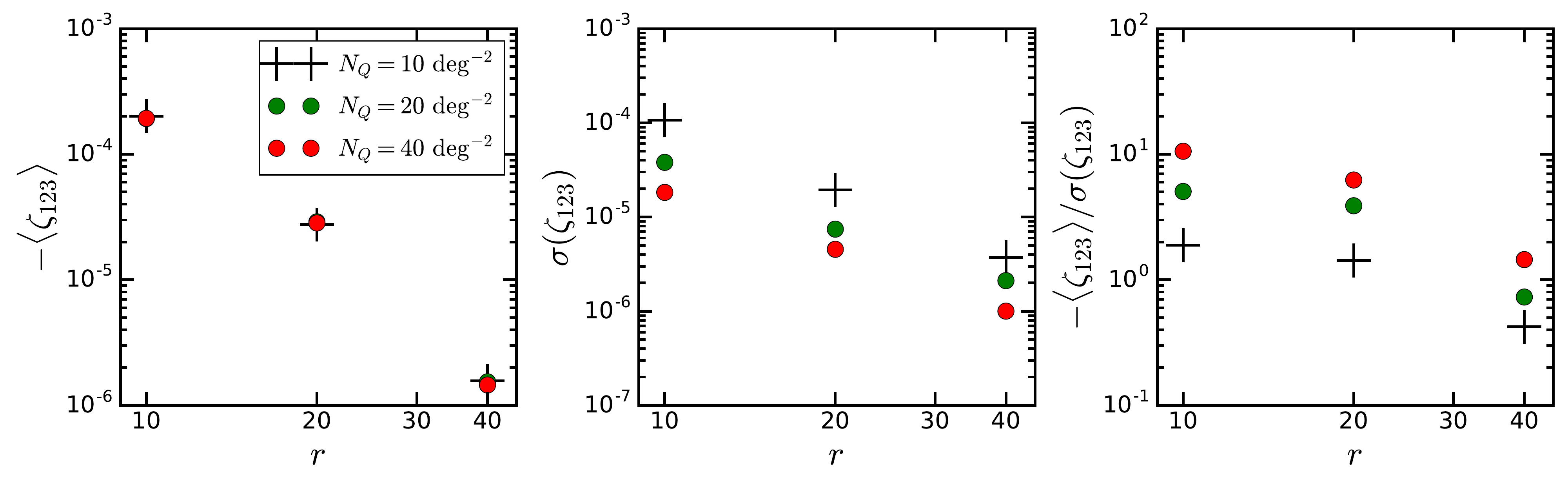}
\caption{Same as Figure \ref{snr}, but assuming different quasar surface densities. We compare only the full resolution spectra for brevity, as the smoothed spectra show similar results.}
\label{snr_nlos}
\end{figure*}

Since we again use the entire 1 $h^{-1}$ Gpc path length, our 4000 and 8000 unique sight-lines effectively result in 12,000 and 24,000 usable \lya\ forest spectra. We show the results in Figure \ref{snr_nlos}. The trend with smoothing the spectra is similar as before, so we only show the comparison at full resolution for brevity. For 8000 sight-lines in our box we expect a SNR of $\sim$ 9, 6, and 2 at $r = 10, 20$, and 40 \hmpc , respectively. To extrapolate our results to DESI, we scale our SNR by $(420,000/24,000)^{1/2} \approx 4$ to get an expected SNR of 37, 25, and 8 at these three separations.

\section{Conclusion}
The standard picture of the Ly$\alpha$ forest is one where the low-density gas in the IGM remains in photoionization equilibrium with the ionizing background and obeys a tight temperature-density relation. Density fluctuations are thought to dominate the structure of the forest, making the Ly$\alpha$ forest a tracer of large-scale structure and a powerful cosmological tool. In principle however, flux fluctuations can arise from other sources such as spatial variations in the IGM mean temperature and the ionizing background.

We have used LyMAS \citep{Peirani2014,Lochhaas2016} to make the first predictions of the 3-d 3PCF fo the \lya\ forest, $\delta_{123} = \langle \delta^F_1 \delta^F_2 \delta^F_3 \rangle$, bootstrapping results from the (100 \hmpc )$^3$ Horizon hydrodynamic simulation \citep{Dubois2014} into a (1 $h^{-1}$ Gpc)$^3$ DM-only simulation. We introduce a simplified ``conditional mean'' formulation of LyMAS, which yields the same results for flux correlation functions as the original ``conditional PDF'' formulation but makes it much easier to implement effects of a fluctuating UVB. To derive a fluctuating radiation field, we assume quasars as our ionizing sources with various radiation mean free paths, and we either randomly distribute them in space or place them in massive DM halos. For our three-point clustering measurements, we focus on triangle configurations with $r_{12} \sim r_{13} \leq 60$ $h^{-1}$ Mpc, and with opening angles $\theta = 90^\circ, 60^\circ,$ and 20$^\circ$. 

The predicted 3PCF of the \lya\ forest approximately follows the hierarchical behavior expected for matter clustering from Gaussian initial conditions: $\zeta_{123} = Q(\xi_{12} \xi_{13} + \xi_{12} \xi_{23} + \xi_{13} \xi_{23})$, with $Q$ approximately independent of scale. For a uniform UVB, we find $Q \approx -4$ to $-5$ on scales of $10-30$ \hmpc , with a weak dependence on triangle size and shape. The large value of $|Q|$ (compared to $|Q| \approx 1$ for matter) likely arises from the low bias factor of the forest, while the negative sign arises because higher densities produce lower fluxes. Even with a (1 $h^{-1}$ Gpc)$^3$ simulation volume, our predictions become noisy on scales larger than $r \sim 30 $ \hmpc . 

For a fluctuating UVB, we consider three values of quasar volume density $n_q$ and three values of the ionizing photon mean free path $\lambda = 300$ \hmpc , 100 \hmpc , and 50 \hmpc\ (comoving). Even the longest of these $\lambda$ is shorter than the observationally inferred value at $z=2.3$, but we would need larger simulation volumes to model larger $\lambda$. Our shorter $\lambda$ values amplify UVB fluctuations to a level expected at higher redshifts; the \cite{Worseck2014} estimate corresponds to a comoving $\lambda \approx 350$ \hmpc\ at $z=3$ and 130 \hmpc\ at $z=4$. 

For randomly-placed quasars, UVB fluctuations boost the 2PCF and 3PCF on all scales, with larger enhancements for smaller $\lambda$ or smaller $n_q$ as expected. With $\lambda = 300$ \hmpc\ the enhancements are small. With $\lambda \leq 100$ \hmpc\ the large scale enhancements are a factor of two or more, making UVB fluctuations the dominant source of large scale flux correlations. The value of $Q$ remains approximately constant on the scales we can reliably measure, with a somewhat smaller $|Q| \sim 3$ for the smaller $\lambda$ values. For halo-based quasars, a fluctuating UVB with $\lambda = 300$ \hmpc\ depresses the 2PCF and 3PCF on all scales relative to a uniform UVB because overdense regions have a higher average UVB that counteracts the higher average IGM density. For $\lambda = 100$ \hmpc\ or $\lambda = 50$ \hmpc\ , the 2PCF and 3PCF are higher than those for a uniform UVB but lower than those for randomly placed quasars. The value of $|Q|$ is nearly unchanged for $\lambda = 300$ \hmpc\ , and it is again moderately reduced (to $|Q| \sim 3$) for $\lambda = 50$ or 100 \hmpc\ . For $\lambda = 300$ \hmpc\ and halo-based quasars, raising $n_q$ from 10$^{-5}$ (\hmpc )$^3$ to 8 $\times$ 10$^{-5}$ (\hmpc )$^3$ further depresses the 2PCF and 3PCF, while lowering $n_q$ to 1.25 $\times$ 10$^{-6}$ (\hmpc )$^3$ strongly boosts the predicted correlation functions and reduces $|Q|$ to $\sim 2$. 

Because hierarchical behavior of the 2PCF and 3PCF is a distinctive prediction of gravitational instability and Gaussian initial conditions, we hoped that we might see a marked transition to a scale-dependent $|Q|$ on scales where UVB fluctuations become a significant driver of flux correlations. However, we do not see a clear sign of such a transition in our results. Unfortunately our simulation volume is too small to yield precise 3PCF measurements on scales $\geq \lambda$. Confirming or refuting the conjecture of a scale-dependent $|Q|$ from UVB fluctuations will require further studies with larger simulation volumes. 

Finally, we derive a rough estimate of the detectability of the 3PCF in data sets such as BOSS, eBOSS, and DESI. We reduce the sight-line density to values comparable to these surveys, consider loosely equilateral triangle configurations that are approximately transverse to the line of sight, and assume that the SNR will scale as $N_q^{-1/2}$, where $N_q$ is the number of quasar sight-lines. In the absence of observational noise, we estimate SNR $\sim$ 7 and $\sim$ 9 for a BOSS- and eBOSS-like data set at $r = 10$ \hmpc\ and 20 \hmpc\ , increasing to $\sim$ 37 and $\sim$ 25 for DESI. At $r = 40$ \hmpc\ the predicted SNR is lower by a factor of $\sim$ 3$-$5. Smoothing or binning the spectra over a scale of 16 BOSS-like pixels barely alters the SNR of the 3PCF measurement, which should simplify observational analyses.

Higher-order moments of large-scale structure contain richer and more complex information than two-point statistics alone. Bispectrum-like measurements along individual lines of sight already show promise as tests of the gravitational instability paradigm for the \lya\ forest and constraints on other sources of structure \citep{Zaldarriaga2001,Fang2004}. Dense, wide-area spectroscopic surveys such as BOSS, eBOSS, and DESI offer the prospect of measuring 3-point correlations of the \lya\ forest in 3-dimensional redshift space. These measurements can provide new diagnostics of non-gravitational physics affecting the \lya\ forest, and reproducing higher-order statistics will allow more confident use of \lya\  forest BAO as a probe of dark energy.

\section*{Acknowledgment}
We thank J. Blaizot, J. Devriendt, Y. Dubois, and C. Pichon for their collaborative contributions to the Horizon hydrodynamic simulation used in this work. We also thank Cassandra Lochhaas for providing her list of dark matter halos and Debopam Som for providing the number of eBOSS \lya\ quasars between $z=2.1$ and $z=2.6$. 

This work was supported by the U.S. Department of Energy, Office of Science, Office of High Energy Physics under Award Number DE-SC-0011726 and NSF grant AST-1516997. It was also supported by collaborative visits funded by the Cosmology and Astroparticle Student and Postdoc Exchange Network (CASPEN). This work has been done in part within the Labex ILP (reference ANR-10-LABX-63) part of the Idex SUPER, and received financial state aid managed by the Agence Nationale de la Recherche, as part of the programme Investissements d'avenir under the reference ANR-11-IDEX-0004-02. TS was funded by Conacyt (Consejo Nacional de Ciencia y Technologia) and UCL (University College London). WZ was supported by the Beatrice and Vincent Tremaine Fellowship. 

This research made use of Astropy, a community-developed core Python package for Astronomy \citep{Astropy2013}.

\section*{Appendix A}
\begin{figure*}
\centering
\includegraphics[width=0.9\textwidth]{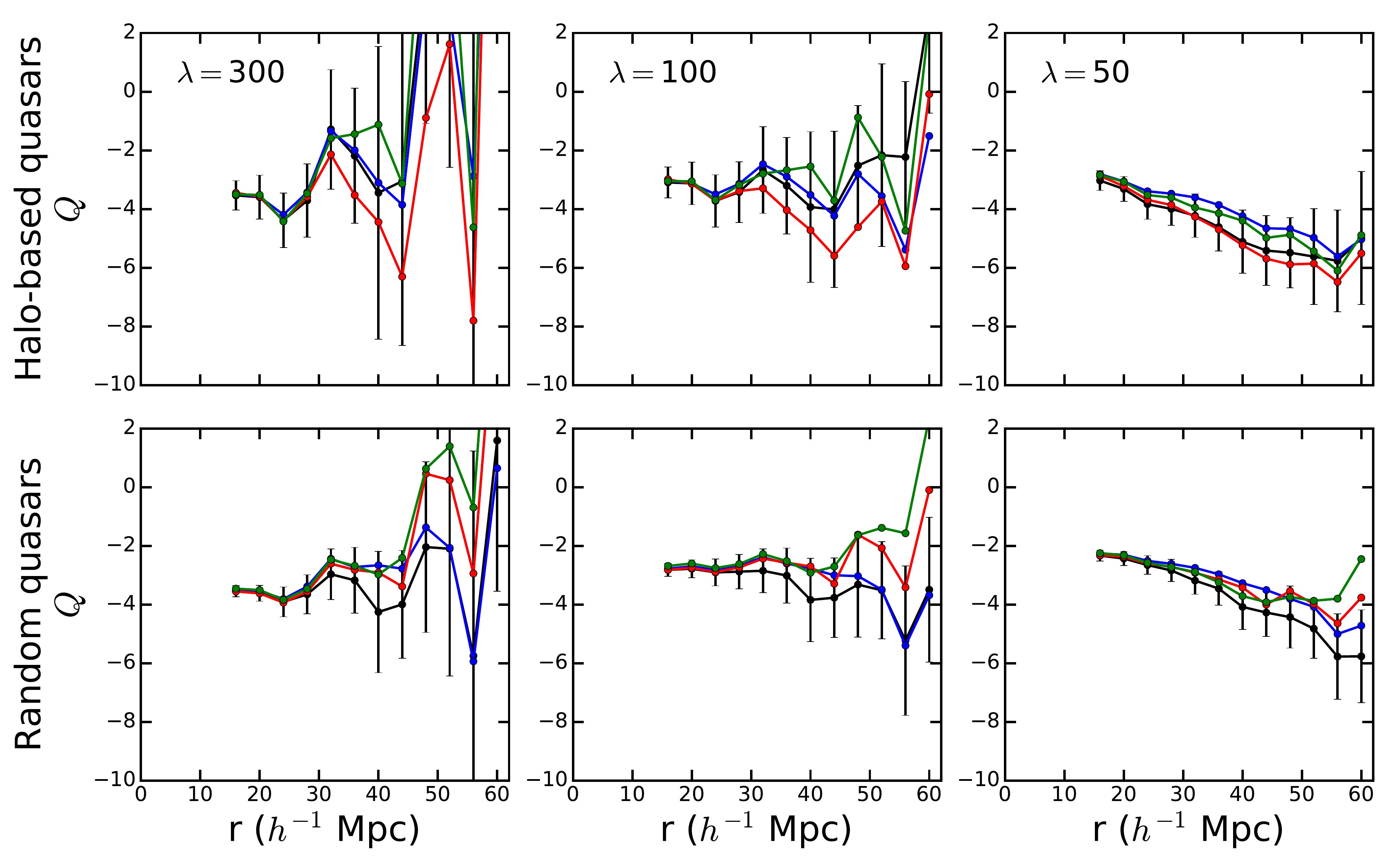}
\caption{The reduced 3PCF from three different random realizations of the quasar population in addition to our fiducial simulation of halo-based and randomly distributed quasars, focusing on triplets with opening angle $\theta = 60^\circ$. Each panel refers to a fluctuating UVB with a different mean free path $\lambda$. Each colored line refers to a realization, with the black line being our fiducial simulation, where we only show the error bars for the fiducial simulation. The variation in the clustering among different realizations of the quasar distribution is within the error bars. The same is true for triplets with $\theta = 90^\circ$ and $\theta = 20^\circ$. }
\label{Q_diffseeds}
\end{figure*}

We check to make sure that our measurements of the correlation functions are stable against different realizations of the quasar distributions. Figure \ref{Q_diffseeds} shows a comparison of the $Q$ amplitude for three extra realizations of halo-based and randomly distributed quasars for the fiducial quasar volume density $n_q = 10^{-5}$ $h^3$ Mpc$^{-3}$. The different realizations display the same overall trend in the reduced 3PCF values in the different UVB, with variations falling within the error bars. 

We next investigate if our clustering measurements are limited by statistics or cosmic variance. We divide our 1 $h^{-1}$ Gpc box with a smooth UV background into nine equal subvolumes and assign each sight-line triplet to a random and the correct subvolume. Assigning triplets to random subvolumes in principle lets us average out the variance due to large scale structure. The correct subvolume assignment is done based on the position of the primary sight-line regardless of the positions of the second and third sight-lines. The random subvolume assignment results in a uniform number of triplets in each subvolume. 

\begin{figure*}
\centering
\includegraphics[width=0.9\textwidth]{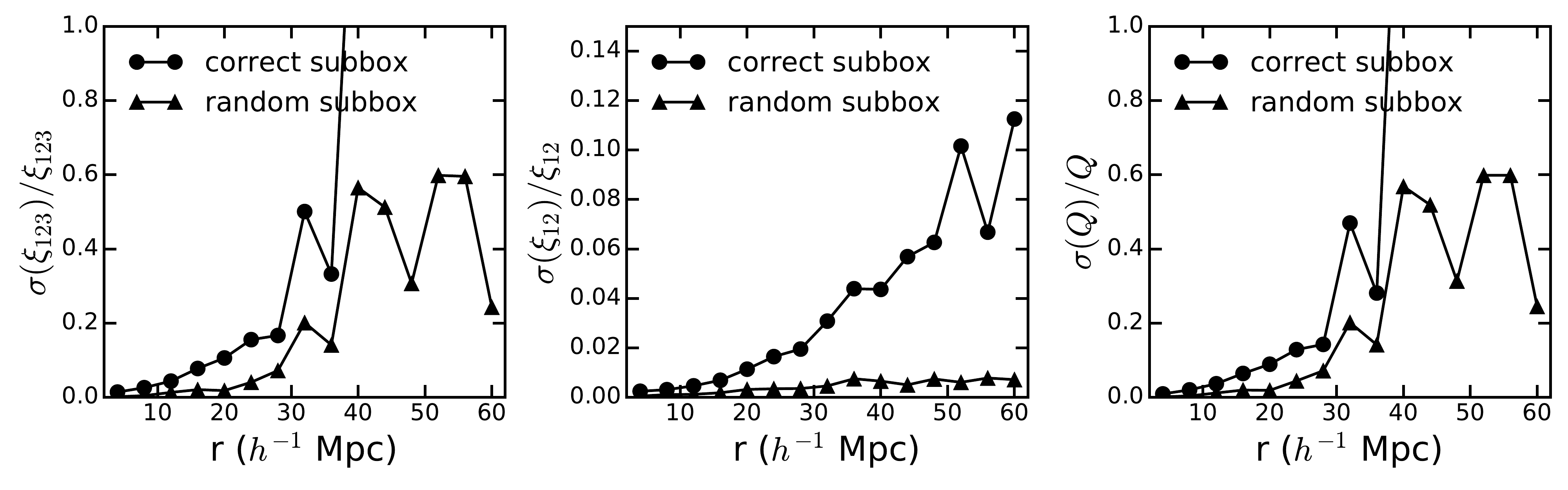}
\caption{Fractional errors in the 3PCF (left), 2PCF (middle), and $Q$ (right) when we assign triplets to the correct (based on the location of the primary sight-line) vs. random subvolumes. Here we use the box with a smooth UV background. Assigning triplets to random subvolumes averages out variance from large scale structure. The fractional errors of the correlation functions are larger for the ``correct subbox'' assignment, therefore suggesting that we are limited by cosmic variance. The errors in $Q$ are dominated by the errors in the 3PCF. }
\label{correct_random_subbox}
\end{figure*}

We then compare the errors bars of the correlation functions measured from triplets using the correct and random subvolume assignment, which is shown in Figure \ref{correct_random_subbox}. We obtain overall larger fractional errors for the 2PCF, 3PCF, and $Q$ when the sight-line triplets are distributed correctly compared to when they are distributed randomly. This is especially true at increasingly large scales, $r \gtrsim$ 10 $h^{-1}$ Mpc. This suggests that our clustering measurements are limited by variance due to large scale structure, rather than by the sampling of these structures from the available sight-lines in our box. 

\begin{figure*}
\centering
\includegraphics[width=0.96\textwidth]{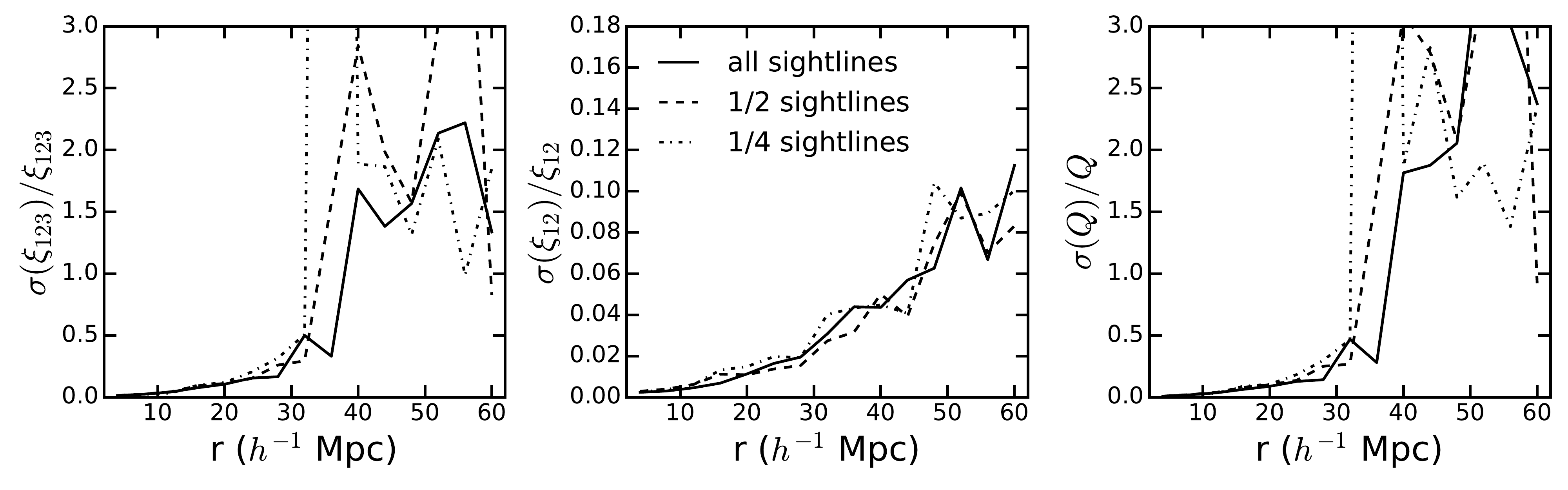}
\caption{Fractional errors in the 3PCF (left), 2PCF (middle), and $Q$ (right) when we vary the number of sight-lines to use in our clustering measurements. There is no significant improvement between using a quarter of or all available sight-lines in the box. This, in combination with Figure~\ref{correct_random_subbox}, suggests that our errors are mostly dominated by cosmic variance rather than by not having enough sight-lines in our box.}
\label{cosmic_variance}
\end{figure*}

We also compare the error bars on the correlation functions from varying the number of sight-lines in our box, in which we use all sight-lines, half of all sight-lines, and a quarter of all sight-lines. As shown in Figure \ref{cosmic_variance}, the fractional errors of the correlation functions are approximately the same whether we use all or a quarter of the available sight-lines. This further suggests that we are not limited by the number of our triplets. 

\end{document}